\documentclass[12pt]{iopart}
\usepackage{graphicx}
\usepackage{dcolumn}
\usepackage{bm}

\usepackage{iopams}
\def\be{\begin{equation}}
\def\ee{\end{equation}}
\def\ba{\begin{array}}
\def\ea{\end{array}}
\def\bea{\begin{eqnarray}}
\def\eea{\end{eqnarray}}
\begin{document}
\title[\underline{J. Phys. G: Nucl. Part. Phys. \hspace {5.9cm} T. R. Routray et al. }]
%{Spin polarised infinite nuclear matter with finite range simple effective interaction and Global description of properties of finite nuclei}
{Study of spin polarized nuclear matter and finite nuclei with finite range simple effective interaction} 

\author{B. Behera$^{1\dagger}$, X. Vi\~nas$^2$, T. R. Routray$^{1*}$ and M. Centelles$^2$}
\address{$^1$School of Physics, Sambalpur University, Jyotivihar-768 019, India.\\
$^2$Departament d'Estructura i Constituents de la Mat\`eria
and Institut de Ci\`encies del Cosmos (ICC),
Facultat de F\'{\i}sica, Universitat de Barcelona,
Diagonal 645, E-08028 Barcelona, Spain
\\
$^*$E-mail: trr1@rediffmail.com (corresponding author)\\
$^{\dagger}$ Retired professor \\ }

\date{\today}

\begin{abstract}
The properties of spin polarized pure neutron matter and symmetric nuclear
matter are studied using the finite range simple effective interaction, upon
its parametrization revisited. Out of the total twelve parameters involved,
we now determine ten of them from nuclear matter, against the nine parameters
in our earlier calculation, as required in order to have predictions in both
spin polarized nuclear matter and finite nuclei in unique manner being free
from uncertainty found using the earlier parametrization. 
The information on the effective mass splitting in polarized
neutron matter of the microscopic calculations is used to constrain the one
more parameter, that was earlier determined from finite nucleus, and in doing
so the quality of the description of finite nuclei is not compromised.
The interaction with the new set of parameters is used to study the
possibilities of ferromagnetic and antiferromagnetic transitions in
completely polarized symmetric nuclear matter. Emphasis is given to analyze
the results analytically, as far as possible, to elucidate the role of the
interaction parameters involved in the predictions. 
\end{abstract}



\bigskip

\section{Introduction}

The study of nuclear matter and finite nuclei in a given model is a subject of
contemporary interest in the area of nuclear research. The most fundamental {\it ab
initio} calculations of Dirac-Brueckner-Hartree-Fock (DBHF), Brueckner-Hartree-Fock (BHF)
and variational types \cite{ter87, muth00, hoffma01,samma10,dbhf3,ebhf05,pand81,Bomb91,Wu07,bal04, akmal01,Wiringa} start from
a Hamiltonian which is adjusted to reproduce the nucleon-nucleon (N-N) scattering
phase shifts and properties of few nucleon bound systems. The predictions in the
regime of nuclear matter (NM) of this kind of {\it ab initio} calculations are usually
considered as a standard. However, the extension to finite nuclei of microscopic
calculations has severe constraints due to the much involved theoretical and
computational procedures. Mean field model calculations
\cite{ring96,brink72,brack85,trr98,ston07} are very popular to deal with finite
nuclei for their relatively simpler computational requirements and analytical
advantages as compared to {\it ab initio} calculations.

Within such a kind of models the Relativistic Mean Field (RMF) model,
which uses an effective Lagrangian constructed by considering meson exchange, 
enjoys an advantageous position for its successful application to structure and
reaction studies of finite nuclei including stable as well as superheavy
nuclei \cite{ring96,lala97,patra01,ddme2,todd05,klahn06,roca11,afan13}.
In the non-relativistic domain, the Skyrme interactions
 \cite{brink72,bein75,rein95,chab97,gor10,erl12}
deserve a similar status for their wide applications to
finite nuclei calculations due to the analytical simplicity
because of the zero-range of these forces. A considerable progress has also been made
in the non-relativistic domain to develop mean field models using finite range
interactions. The Gogny \cite{gogny80,blaizot95,chappert08,goriely09}, M3Y
\cite{nakada03,than09} and the finite range simple effective interaction (SEI) 
\cite{trr98,trr02,trr13} forces are 
examples of this type of interactions. In addition, it is worth mentioning the recent development of finite range functionals using the effective field theory methodology applied to low-energy nuclear physics \cite{raim14}. 

The equation of state (EOS) and the momentum dependence of the mean field are two
important aspects in the studies of NM (a compilation of several acronyms used in this paper is provided in Table 1). However, effective mean field models may 
predict results in NM that do not necessarily agree with the results provided by
microscopic calculations.
For example, microscopic calculations predict a larger effective mass for neutrons 
than for protons in neutron-rich isospin-asymmetric nuclear matter (ANM) \cite{ebhf05,ma04,sammu05,zuo05}, which is
also the currently accepted point of view, as substantiated by the
experimental results of the energy dependence of the Lane potential
\cite{lane62,hod94}. However, mean field calculations based on effective interactions 
in both the relativistic and the non-relativistic  
domains do not always fulfill this trend 
\cite{kubis97,greco03,dutra12,behera05}. 
This points out to the fact
that the momentum dependence of the mean field in NM has not been taken as a
prerequisite in the fitting of the parameters of the RMF and  most of the effective non-relativistic models. 
As a consequence, there is no reason {\it a priori} why these models should follow the
trend of momentum dependent properties, such as mean field, effective mass
splitting, etc. as provided by microscopic calculations in NM. The parameters
of effective relativistic and non-relativistic models are usually
constrained from the empirical values of some NM properties, such as energy per
particle $e(\rho)$, incompressibility $K(\rho)$, symmetry energy $E_{s}(\rho)$,
etc., computed at saturation density $\rho$=$\rho_0$ and to some selected
experimental data of binding energies (BE) and radii over the periodic table. The
empirical values of $\rho_0$, $e(\rho_0)$ and $E_s(\rho_0)$ used in these models
vary within the ranges 0.17$\pm$0.03 fm$^{-3}$, $-16\pm$0.2 MeV and 33$\pm$5 MeV,
respectively. With fitting protocols of this type, the momentum dependence of the
mean field is completely predicted by the model and can or cannot reproduce the
tendency exhibited by the microscopic calculations.

\begin{table*} \caption{Several acronyms used in the text.}
\begin{center}
\begin{tabular}{ll}\hline \hline
Acronym & Meaning \\ \hline
SEI   & Simple effective interaction \\
NM    & Nuclear matter \\
BE    & Binding energy \\
SNM   & Symmetric nuclear matter \\
ANM   & Asymmetric nuclear matter \\
PNM   & Pure neutron matter \\
PPNM  & Polarized pure neutron matter \\
CPNM  & Completely polarized pure neutron matter \\
FM    & Ferromagnetic \\
AFM   & Antiferromagnetic \\
CSNM  & Ferromagnetic completely polarized SNM \\
CASNM & Antiferromagnetic completely polarized SNM \\
\hline \hline
\end{tabular}
\end{center}
\end{table*}

The momentum dependence of the mean field is a fundamental property
\cite{bers88,welke88} and it should not be left open to assume an arbitrary
behaviour. This momentum dependence of the mean field, as extracted from the
analysis of nucleon-nucleus scattering data \cite{welke88,gale90,cser92,Danielz00}
is explicitly taken into account in the fitting procedure of the parameters of the
finite range simple effective interaction (SEI) which has been used in NM studies
\cite{behera05,trr05,trr07,trr09,trr11}. 
In a recent work \cite{trr13}, the SEI has
been extended for studying ground-state finite nuclei properties. It should be
pointed out that the SEI depends on eleven parameters (apart from a spin-orbit strength
parameter in the case of finite nuclei) of which nine could be obtained from the
studies of ANM. The parameters responsible for the momentum dependence of the
nucleonic mean fields in ANM are decided by utilizing appropriate constraints with
care that these predictions are not changed while deciding the rest of the
parameters. The two open parameters of SEI, $t_0$ and $x_0$, and the spin-orbit strength $W_0$ are left to reproduce a few magic nuclei.
Within this protocol to determine the parameters of the SEI, the microscopic trends of the properties in ANM could be reproduced and, at the same time, the known binding energies and charge radii of even-even spherical nuclei could be described with a quality similar to other successful mean field models of relativistic or non-relativistic type (see table 3 of \cite{baldo08}) \cite{trr13}.
However, while examining the results in spin polarized NM it is found that the determination of the parameters $t_0$, $x_0$ and $W_0$ from finite nuclei does not yield unique predictions in spin polarized NM. In this work our objective is to study the
spin polarized NM using the SEI in consonance with the microscopic predictions
to remove the aforementioned ambiguity.
This is done by using the microscopic information on the momentum
dependence of the mean field in spin polarized pure neutron matter (PPNM)
to fix the $x_0$ parameter, leaving $t_0$ and $W_0$ only to be
determined exclusively from finite nuclei. Our procedure enables to constrain the
strengths of the interaction in the four basic channels of the $N-N$ interaction,
namely, the singlet-even $V^{SE}$, triplet-even $V^{TE}$, triplet-odd
$V^{TO}$ and singlet-odd $V^{SO}$ channels as well as the predictions in spin
polarized NM, in a unique manner for a given EOS. In this new method 
of determining the parameters,
the earlier predictions in ANM do not change and the finite nuclei
results are also reproduced within reasonable accuracy. Thus the obtained SEI can be
used in the study of both isospin and spin polarized NMs as well as in finite nuclei.
In section 2 we outline the formulation of spin polarisation in
symmetric nuclear matter (SNM) and pure neutron matter (PNM) using the SEI. 
In this section we also discuss the determination of the parameters from ANM and finite nuclei,
as well as the underlying uncertainty that manifests in the predictions in the spin channel. 
In section 3 the procedure adopted to remove the uncertainty
is worked out and the new SEI parameter sets for EOSs having different
NM incompressibility are obtained. The predictions in the spin channel in
PNM and SNM are discussed and compared with the results of other mean field
models as well as with microscopic calculations. Section 4
contains a brief summary and  conclusions of our analysis.

\section{Formalism}

The finite range simple effective interaction (SEI) used in the present work is given as
\begin{eqnarray}
v_{eff}(r)&=&t_0 (1+x_0P_{\sigma})\delta(r) \nonumber \\
&&+\frac{t_3}{6}(1+x_3 P_{\sigma})\left(\frac{\rho({\bf R})}
{1+b\rho({\bf R})}\right)^{\gamma} \delta(r) \nonumber \\
&& + \left(W+BP_{\sigma}-HP_{\tau}-MP_{\sigma}P_{\tau}\right)f(r),
\label{eq1}
\end{eqnarray}
where, $f(r)$ is the functional form of the finite range interaction containing the
single range parameter $\alpha$ and is taken to be of Gaussian form,
$e^{-r^2/\alpha^2}$. The other terms have their usual meaning. The SEI in equation
(\ref{eq1}) has 11 parameters, namely, $b$, $t_0$, $x_0$, $t_3$, $x_3$, $\gamma$, $\alpha$,
$W$, $B$, $H$ and $M$ (the spin-orbit strength parameter $W_0$ will enter in 
the formulation of finite nuclei). The complete study of ANM requires the knowledge of altogether
nine parameters, namely, $b$, $\gamma$, $\alpha$, $\varepsilon_{ex}^{l}$,
$\varepsilon_{ex}^{ul}$, $\varepsilon_{\gamma}^{l}$, $\varepsilon_{\gamma}^{ul}$,
$\varepsilon_{0}^{l}$ and $\varepsilon_{0}^{ul}$, with, the indices "l" and "ul"
depicting interactions between pairs of isospin-like and unlike nucleons,
respectively \cite{trr07,trr09}. For the sake of simplicity, the formulation 
has been based on the fact that the range between a 
pair of isospin-like or unlike nucleons is the same but they differ in their
strengths. The connection between the parameters of ANM and the interaction
parameters is given in the earlier works \cite{trr13,trr07}. Here we shall
write them in terms of the strengths of the finite range part of the N-N interaction
in the four states, namely, singlet-even $V_0^{SE}$, triplet-even $V_0^{TE}$,
triplet-odd $V_0^{TO}$ and singlet-odd $V_0^{SO}$ instead of $W$, $B$, $H$ and $M$ for
the sake of convenience of the discussions in this work. They read as,
\begin{eqnarray}
&&\varepsilon_{0}^{ul}=\frac{t_0}{2}\rho_0\left(2+x_0\right)
%\nonumber \\&+&
+ \frac{\rho_0}{8} \left(V_0^{SE}+3V_0^{TE}+3V_0^{TO}+V_0^{SO}\right)
\int f(r)d^3r
\label{eq2.a}
\end{eqnarray}
\begin{eqnarray}
&&\varepsilon_{0}^{l}=\frac{t_0}{2}\rho_0\left(1-x_0\right)
%\nonumber \\&+&
+ \frac{\rho_0}{4}\left(V_0^{SE}+3V_0^{TO}\right)\int f(r)d^3r 
\label{eq2.b}
\end{eqnarray}
\begin{eqnarray}
\varepsilon_{\gamma}^{ul}&=&\frac{t_3}{12}\rho_0^{\gamma+1}(2+x_3)
\label{eq2.d}
\end{eqnarray}
\begin{eqnarray}
\varepsilon_{\gamma}^{l}&=&\frac{t_3}{12}\rho_0^{\gamma+1}(1-x_3)
\label{eq2.c}
\end{eqnarray}
\begin{eqnarray}
\varepsilon_{ex}^{ul}&=&\frac{\rho_0}{8}\left(V_0^{SE}+3V_0^{TE}-3V_0^{TO}
-V_0^{SO}\right)\int f(r)d^3r
\label{eq2.e}
\end{eqnarray}
\begin{eqnarray}
\varepsilon_{ex}^{l}&=&\frac{\rho_0}{4}\left(V_0^{SE}-3V_0^{TO}\right)
\int f(r)d^3r .
\label{eq2.f} 
\end{eqnarray}
%For our Gaussian form of $f(r)$, $\int f(r)d^3r=\pi^{3/2}\alpha^3$. 
The knowledge of the nine parameters that characterize the ANM can be obtained from 
the independent studies of spin saturated SNM and PNM. 

In SNM the numbers of neutrons and protons are equal. In the case of completely polarized SNM, there are two possibilities for the spin polarisation.
One of them corresponds to the situation where the spins of neutrons and protons are aligned in the same direction (CSNM), referred to as ferromagnetic (FM) type. The other possibility corresponds to the situation where the spins of neutrons and protons are oriented in opposite directions (CASNM), referred to as anti-ferromagnetic (AFM) type. For the SEI, the energy per particle (obtained as the ratio of the energy density $H$ with the density $\rho$) in SNM, CSNM, and CASNM is given, respectively, by 
\begin{eqnarray} %\nonumber
e(\rho) &=& \frac {H(\rho)}{\rho} = \frac{3\hslash^2k_f^2}{10M}
+\frac{(\varepsilon_{0}^{l}+\varepsilon_{0}^{ul})}{4\rho_0}\rho
% \\ &&
+\frac{(\varepsilon_{\gamma}^{l}+\varepsilon_{\gamma}^{ul})}{4\rho_0^{\gamma+1}}
\rho\left(\frac{\rho({\bf R})}
{1+b\rho({\bf R})}\right)^{\gamma}
\nonumber \\
&& +\frac{(\varepsilon_{ex}^{l}+\varepsilon_{ex}^{ul})}{4\rho_0}
\rho J(k_f)
%\bigg[\frac{3\Lambda^6}{16k_f^6}-\frac{9\Lambda^4}{8k_f^4} \nonumber \\
%&&+\left(\frac{3\Lambda^4}{8k_f^4}-\frac{3\Lambda^6}{16k_f^6}\right)
%e^{-4k_f^2/\Lambda^2} \nonumber \\ 
%&& + \frac{3\Lambda^3}{2k_f^3}\int_0^{2k_f/\Lambda}e^{-t^2}dt \bigg] ,
\label{eq8}
\end{eqnarray}
\begin{eqnarray} %\nonumber
e_{pol}^{S}(\rho) &=& \frac {H_{pol}^{S}(\rho)}{\rho} =
\frac{3\hslash^2{k_f^{pol}}^2}{10M}
+\frac{\varepsilon_{0}^{ls}}{2\rho_0}\rho
% \\ \nonumber &+&
+ \frac{\varepsilon_{\gamma}^{ls}}{2\rho_0^{\gamma+1}}
\rho\left(\frac{\rho({\bf R})} {1+b\rho({\bf R})}\right)^{\gamma} \nonumber \\
&+&\frac{\varepsilon_{ex}^{ls}}{2\rho_0}
\rho J(k_{f}^{pol})
%\bigg[\frac{3 \Lambda^6}{16{k_f^{pol}}^{6}}
%-\frac{9 \Lambda^4}{8{k_f^{pol}}^{4}} \nonumber \\
%&+&\left(\frac{3\Lambda^4}{8{k_f^{pol}}^{4}}-\frac{3\Lambda^6}{16{k_f^{pol}}^{6}}\right)
%e^{-4{k_f^{pol}}^{2}/\Lambda^2} \nonumber \\
%&+&\frac{3\Lambda^3}{2{k_f^{pol}}^{3}}\int_0^{2k_f^{pol}/\Lambda}
%e^{-t^2}dt \bigg] 
\label{eq9}
\end{eqnarray}
\begin{eqnarray} %\nonumber
e_{pol}^{AS}(\rho) &=& \frac {H_{pol}^{AS}(\rho)}{\rho} =
\frac{3\hslash^2{k_f^{pol}}^2}{10M}
+\frac{\varepsilon_{0}^{las}}{2\rho_0}\rho
% \\ \nonumber &+&
+ \frac{\varepsilon_{\gamma}^{las}}{2\rho_0^{\gamma+1}}
\rho\left(\frac{\rho({\bf R})} {1+b\rho({\bf R})}\right)^{\gamma} \nonumber \\
&+&\frac{\varepsilon_{ex}^{las}}{2\rho_0}
\rho J(k_{f}^{pol}) ,
%\bigg[\frac{3 \Lambda^6}{16{k_f^{pol}}^{6}}
%-\frac{9 \Lambda^4}{8{k_f^{pol}}^{4}} \nonumber \\
%&+&\left(\frac{3\Lambda^4}{8{k_f^{pol}}^{4}}-\frac{3\Lambda^6}{16{k_f^{pol}}^{6}}\right)
%e^{-4{k_f^{pol}}^{2}/\Lambda^2} \nonumber \\
%&+&\frac{3\Lambda^3}{2{k_f^{pol}}^{3}}\int_0^{2k_f^{pol}/\Lambda}
%e^{-t^2}dt \bigg] ,
\label{eq10}
\end{eqnarray}
where
\begin{eqnarray}
J(k_i)= \frac{3\Lambda^3}{2{k_i^3}} \bigg[\frac{\Lambda^3}{8{k_i^3}}
-\frac{3 \Lambda}{4{k_i}}
 -\left(\frac{\Lambda^3}{8{k_i^3}}-\frac{\Lambda}{4{k_i}}\right) e^{-4{k_i^2}/\Lambda^2}
+\frac{\sqrt{\pi}}{2} erf\left(2k_i/\Lambda\right) \bigg]
%+\int_0^{2k_i/\Lambda}\!\! e^{-t^2}dt \bigg]
\label{eq10a}
\end{eqnarray}
and $\Lambda$=$ \frac {2}{\alpha}$. The expressions 
of $J(k_i)$ in equations (\ref {eq8}), (\ref {eq9}) and (\ref {eq10}) 
can be obtained from equation (\ref {eq10a}) by using 
$k_i = k_f = (\frac{3}{2}{\pi}^{2}\rho)^{\frac{1}{3}}$  and 
$k_i = k_f^{pol}  = (3{\pi}^{2}\rho)^{\frac{1}{3}}$, where $k_f$ is the Fermi momentum
in SNM and $k_f^{pol}$ is the Fermi momentum in CSNM
and CASNM. The new parameters appearing in equations
(\ref{eq9}) and (\ref{eq10}) with index ``ls" and ``las" are given as, 
\begin{eqnarray}
&&\varepsilon_{0}^{ls}=\frac{t_0}{2}\rho_0\left(1+x_0\right)
%\nonumber \\&+&
+ \frac{\rho_0}{4}\left(V_0^{TE}+3V_0^{TO}\right)\int f(r)d^3r 
\label{eq11a}
\end{eqnarray}
\begin{eqnarray}
\varepsilon_{\gamma}^{ls}&=&\frac{t_3}{12}\rho_0^{\gamma+1}(1+x_3)
\label{eq12a}
\end{eqnarray}
\begin{eqnarray}
\varepsilon_{ex}^{ls}&=&\frac{\rho_0}{4}\left(V_0^{TE}-3V_0^{TO}\right)
\int f(r)d^3r,
\label{eq13a} 
\end{eqnarray}
\begin{eqnarray}
&&\varepsilon_{0}^{las}=\frac{t_0}{2}\rho_0
%\nonumber \\&+&
+ \frac{\rho_0}{8}\left(V_0^{TE}+V_0^{SE}+5V_0^{TO}+V_0^{SO}\right)\int f(r)d^3r 
\label{eq11b}
\end{eqnarray}
\begin{eqnarray}
\varepsilon_{\gamma}^{las}&=&\frac{t_3}{12}\rho_0^{\gamma+1}
\label{eq12b}
\end{eqnarray}
\begin{eqnarray}
\varepsilon_{ex}^{las}&=&\frac{\rho_0}{4}\left(V_0^{TE}+V_0^{SE}-5V_0^{TO}-V_0^{SO}\right)\int f(r)d^3r.
\label{eq13b} 
\end{eqnarray}
Similarly, the energy per particle in PNM, PPNM and completely polarized PNM 
(CPNM) can be given for the SEI as, 
\begin{eqnarray} \nonumber 
e^N(\rho) &=& \frac{H^N(\rho)}{\rho} = \frac{3\hslash^2k_n^2}{10M}
+\frac{\varepsilon_{0}^{l}}{2\rho_0}\rho \nonumber \\
&+&\frac{\varepsilon_{\gamma}^{l}}{2\rho_0^{\gamma+1}}
\rho\left(\frac{\rho({\bf R})} {1+b\rho({\bf R})}\right)^{\gamma}
+\frac{\varepsilon_{ex}^{l}}{2\rho_0}
\rho J(k_n)
\label{eq14}
\end{eqnarray}
\begin{eqnarray} \nonumber
e_{pol}^{N}(\rho,\beta_{\sigma}) &=&
\frac{H_{pol}^N(\rho_{nu},\rho_{nd})}{\rho} =
\frac{1}{\rho}\bigg[ \frac{3\hslash^{2}(k_{nu}^{2}\rho_{nu}+k_{nd}^{2}\rho_{nd})}{10M}
+\frac{\varepsilon_{0}^{l,l}}{2\rho_0}(\rho_{nu}^{2}
+\rho_{nd}^{2}) \nonumber \\
&+&\frac{\varepsilon_{0}^{l,ul}}{\rho_0}\rho_{nu}\rho_{nd}
+ \bigg(\frac{\varepsilon_{\gamma}^{l,l}}{2\rho_0^{\gamma+1}}
(\rho_{nu}^{2}+\rho_{nd}^{2})+\frac{\varepsilon_{\gamma}^{l,ul}}
{\rho_0^{\gamma+1}}\rho_{nu}\rho_{nd} \bigg)\left(\frac{\rho({\bf R})}
{1+b\rho({\bf R})}\right)^{\gamma} \nonumber \\
&+&\frac{\varepsilon_{ex}^{l,l}}{2\rho_0} \bigg( \rho_{nu}^{2} J(k_{nu})
+\rho_{nd}^{2} J(k_{nd}) \bigg) \nonumber \\
&+&\frac{\varepsilon_{ex}^{l,ul}}{4\rho_0\pi^{2}} \bigg(
\rho_{nu} \int_0^{k_{nd}}\!\! I(k,k_{nu})k^{2}dk
+ \rho_{nd} \int_0^{k_{nu}}\!\! I(k,k_{nd})k^{2}dk \bigg) \bigg]
\nonumber \\
\label{ppnm}
\end{eqnarray}
and
\begin{eqnarray} \nonumber 
e_{cpnm}^{N}(\rho) &=& \frac {H_{cpnm}^{N}(\rho)}{\rho} =
\frac{3\hslash^2{k_n^{pol}}^2}{10M}
+\frac{\varepsilon_{0}^{l,l}}{2\rho_0}\rho \\
&+&\frac{\varepsilon_{\gamma}^{l,l}}{2\rho_0^{\gamma+1}}
\rho\left(\frac{\rho({\bf R})}
{1+b\rho({\bf R})}\right)^{\gamma}
+\frac{\varepsilon_{ex}^{l,l}}{2\rho_0}
\rho J(k_{n}^{pol}),
\label{eq15}
\end{eqnarray}
where, in equation (\ref{ppnm}),
\begin{eqnarray}
I(k,k_{i}) &=& \frac{3 \Lambda^3}{8k_{i}^{3}}
\bigg[\frac{\Lambda}{k}
\left(e^{-\left(\frac{k+k_{i}}{\Lambda}\right)^2}
-e^{-\left(\frac{k-k_{i}}{\Lambda}\right)^2}\right) \nonumber \\
&+&
\sqrt{\pi}\left(\textrm{erf}{\bigg(\frac{k+k_{i}}
{\Lambda}\bigg)}-\textrm{erf}{\bigg(\frac{k-k_{i}}
{\Lambda}\bigg)}\right) \bigg]
%+ 2 \int_{\left(\frac{k-k_{i}}
%{\Lambda}\right)}^{\left(\frac{k+k_{i}}
%{\Lambda}\right)}e^{-t^2}dt\bigg] 
\label{ppnma}
\end{eqnarray}
for $k_{i}=k_{nu}$, $k_{nd}$.
One has $k_n$=$(3{\pi}^{2}\rho)^{\frac{1}{3}}$ for the Fermi momentum in PNM, and
$k_{nu(nd)}$=$(6{\pi}^{2}\rho_{nu(nd)})^{\frac{1}{3}}$ and $k_{n}^{pol}$=
$(6{\pi}^{2}\rho)^{\frac{1}{3}}$ for the Fermi momentum in PPNM and CPNM, respectively.
The expressions
of $J(k_i)$ in equations (\ref{eq14}), (\ref{ppnm}) and (\ref{eq15})
can be obtained from equation (\ref{eq10a}) with the use of the respective Fermi
momentum in place of $k_i$. The indices ``$l,l$" and ``$l,ul$" are used to denote the
interaction between a pair of neutrons having the same and opposite spin
orientations, respectively. The splitting of the strength parameters
$\varepsilon_{0}^{l}$, $\varepsilon_{\gamma}^{l}$ and
$\varepsilon_{ex}^{l}$ of PNM are subject to the condition that
$\varepsilon_{0}^{l}$=$(\varepsilon_{0}^{l,l}+\varepsilon_{0}^{l,ul})$/2,
$\varepsilon_{\gamma}^{l}$=$(\varepsilon_{\gamma}^{l,l}+\varepsilon_{\gamma}^{l,ul})$/2
and $\varepsilon_{ex}^{l}$=$(\varepsilon_{ex}^{l,l}+\varepsilon_{ex}^{l,ul})/2$. The exchange strength parameter
${\varepsilon_{ex}^{l,l}}$ in CPNM in equation (\ref{eq15}) can be expressed in terms of the finite range strength in the $TO$ state of the N-N interaction as,
\begin{eqnarray}
{\varepsilon_{ex}^{l,l}} = -\rho_0V_0^{TO}\int f(r)d^3r.
\label{eq16}
\end{eqnarray}
The other parameters of CPNM for SEI in equation (\ref{eq15}) are
${\varepsilon_{\gamma}^{l,l}}$=0 (due to the zero range of the density-dependent term of SEI) and ${\varepsilon_{0}^{l,l}}=-{\varepsilon_{ex}^{l,l}}$. 
The SNM is completely determined by the parameters $b$, $\gamma$,
$\alpha$ and the combinations
%$\frac{(\varepsilon_{0}^{l}+\varepsilon_{0}^{ul})}{2\rho_0}$,
%$\frac{(\varepsilon_{\gamma}^{l}+\varepsilon_{\gamma}^{ul})}{2\rho_0}$ and
%$\frac{(\varepsilon_{ex}^{l}+\varepsilon_{ex}^{ul})}{2\rho_0}$ which are
%combinations of the strength parameters of isospinwise-like and unlike pairs
%in ANM. These combined strength parameters in SNM are represented by the
%symbols,
\begin{eqnarray}
\left(\frac{\varepsilon_{0}^{l}+\varepsilon_{0}^{ul}}{2}\right)=\varepsilon_0;
\left(\frac{\varepsilon_{\gamma}^{l}+\varepsilon_{\gamma}^{ul}}{2}\right)=\varepsilon_{\gamma};
\left(\frac{\varepsilon_{ex}^{l}+\varepsilon_{ex}^{ul}}{2}\right)=\varepsilon_{ex}.
\label{eq17}
\end{eqnarray}
These strength parameters in SNM can also be written as,
%%equations
\begin{eqnarray}
\varepsilon_{ex}={\frac{\rho_0}{16}}(3V_{0}^{SE}+3V_{0}^{TE}-9V_{0}^{TO}
-V_{0}^{SO})\int f(r)d^3r, 
%{\pi^{3/2}}{\alpha^{3}},
\label{eq18}
\end{eqnarray}
\begin{eqnarray}
\varepsilon_{0}&=&{\frac{3}{4}}t_0\rho_0+{\frac{\rho_0}{16}}(3V_{0}^{SE} 
%\nonumber \\&+&
+3V_{0}^{TE}+9V_{0}^{TO}+V_{0}^{SO})\int f(r)d^3r, 
%{\pi^{3/2}}{\alpha^{3}}, 
\label{eq19}
\end{eqnarray}
\begin{eqnarray}
{\varepsilon_{\gamma}}={\frac{t_3}{8}}\rho_0^{\gamma+1}.
\label{eq20}
\end{eqnarray} 
In the foregoing equations, $\int f(r)d^3r=\pi^{3/2}\alpha^3$, can be replaced
where ever it ocures for the Gaussian form of $f(r)$. 

We shall now briefly outline the procedure of determination of the parameters, as
adopted in previous studies of NM and finite nuclei \cite{trr13}. The range $\alpha$
and the exchange strength $\varepsilon_{ex}$ in SNM are determined by means of a
simultaneous minimization procedure using the experimentally extracted constraint
\cite{bers88,welke88,Danielz00} that the attractive optical potential changes sign
for a kinetic energy 300 MeV of the incident nucleon. The NM values of the
saturation density $\rho_0$ and energy per particle $e(\rho_0)$ at saturation are
the only quantities needed to completely determine $\alpha$ and $\varepsilon_{ex}$
(see \cite{trr98} for details). The parameter $b$ is fixed for avoiding the
supraluminous behaviour in SNM \cite{trr97}.
It reads $b\rho_0$=$ \left[ \left(\frac {Mc^2}{T_{f_0}/5-e(\rho_0)} 
\right)^{\frac {1}{ \left(\gamma+1 \right)}}-1 \right]^{-1}$,
with $T_{f_0}$=$\frac{\hslash^2k_{f_0}^2}{2M}$ where $k_{f_0}$ is the Fermi momentum in SNM at normal density and $M$ is the nucleonic mass. Its
calculation requires again the knowledge of the NM values $\rho_0$, $e(\rho_0)$ and the
parameter $\gamma$. The stiffness parameter $\gamma$ determines the density
dependence of the EOS in SNM. The two remaining parameters in SNM, namely
$\varepsilon_{\gamma}$ and $\varepsilon_{0}$, are obtained from the saturation
conditions, that is, from the values of $e(\rho_0)$ and $\rho_0$. The stiffness
parameter $\gamma$ is kept open and its admissible values are constrained by the
condition that the pressure-density curve lies within the region extracted from the
analysis of flow data in heavy-ion collisions (HIC) at intermediate and high
energies \cite{Danielz02}. Thus, a complete study of SNM can be performed for a
given $\gamma$ if one assumes standard NM values of $\rho_0$ and $e(\rho_0)$.

To extend the study for ANM, one needs to know how 
$\varepsilon_{ex}$, $\varepsilon_{\gamma}$ and $\varepsilon_{0}$ split into like and
unlike isospin channels. The splitting of $\varepsilon_{ex}$ into
$\varepsilon_{ex}^{l}$ and $\varepsilon_{ex}^{ul}$ is decided using the physical
constraint resulting from the studies of the thermal evolution of NM properties
\cite{trr11}. This study predicts a critical value of the splitting of
$\varepsilon_{ex}$  for which the thermal evolution of NM properties as well as the
entropy  per particle in PNM does not exceed that of SNM. The resulting critical
value  of the splitting is $\varepsilon_{ex}^{l}$ = $\frac{2}{3}\varepsilon_{ex}$.
The n-p effective mass splitting predicted with this choice of
$\varepsilon_{ex}^{l}$  nicely coincides with the results of DBHF calculations
\cite{samma10} as has been shown in the previous work \cite{trr13}. The splitting of
the remaining two parameters, namely $\varepsilon_{\gamma}$ and $\varepsilon_{0}$,
is obtained by assuming, on the one hand, a standard value of $E_s(\rho_0)$ at
saturation and, on the other hand, the value of its derivative
$E_s^{'}(\rho_0)$ = $\rho_0 \frac{dE_s(\rho_0)}{d\rho_0}$ for which the asymmetric
contribution of the nucleonic part of the energy density in charge neutral
beta-stable $n+p+e+\mu$ matter (referred to as neutron star matter) becomes maximum.
This choice predicts a density dependence of the symmetry energy which is neither
very stiff nor soft and does not allow the direct URCA process to occur in neutron
stars. The population synthesis models \cite{pop06} based on cooling calculations
\cite{blas04} predict that there shall be no direct URCA process at least
in typical neutron stars. Constraining the splitting of the three strength
parameters $\varepsilon_{ex}$, $\varepsilon_{\gamma}$ and $\varepsilon_{0}$ allows
one to determine the nine parameters that describe the ANM. The SEI with the
parameters obtained in this way has the ability of reproducing the microscopic
trends of the density dependence of the EOS and the momentum dependence of the mean
fields in ANM \cite{trr07,trr09}.

There are still two parameters open, which were taken to be $t_0$ and $x_0$
in the previous work \cite{trr13} and were determined from finite nuclei
calculations. The energy density of a finite nucleus was constructed from
the nuclear, Coulomb and spin-orbit
interactions. An improved semi-classical $\hslash^2$-approximation
\cite{vinas00,vinas03} was used to localize the exchange contributions of the nuclear
part. The energy density was thus expressed in terms of the local variables, namely,
nucleon densities, kinetic energy densities and spin-densities. Utilising the
variational principle results into Skyrme-like Hartree-Fock equations which were
solved to get the neutron and proton orbitals. Using this energy density functional,
directly derived from the SEI, one could determine the two pending parameters $t_0$
and $x_0$ along with $W_0$, the strength of the spin-orbit interaction, from the
experimental binding energy of the double-closed-shell nuclei $^{40}$Ca and
$^{208}$Pb and from the $1p_{3/2}$--$1p_{1/2}$ level splitting in $^{16}$O as
discussed in Ref.\cite{trr13}. In the calculation of the ground-state properties of
open-shell nuclei, the pairing correlations were considered in the BCS approach
with a density dependent zero-range pairing interaction \cite{ber91}. The binding energies and charge radii of magic nuclei
reproduced the corresponding experimental values within an accuracy of 0.1${\%}$.
It was also found that the energy density functional associated to the SEI including
pairing was well suited for describing binding energies and charge radii of open shell
nuclei. The experimental binding energies of 161 even-even spherical nuclei and the
measured charge radii of 86 even-even spherical nuclei from $^{16}$Ne to $^{224}$U
were reproduced by our model within windows of $\pm$2 MeV and $\pm$0.02 fm with
overall root mean square ({\it rms}) deviations of 1.54 MeV and 0.015 fm, respectively. These deviations
are in consonance with those obtained in other common mean field interactions. For
example, as given in Table 3 of Ref.\cite{baldo08}, the corresponding {\it rms}
deviations in binding energies and charge radii for the same set of nuclei considered
here are of 1.71 MeV and 0.024 fm in the Skyrme force SLy4, of 3.58 MeV and 0.020 fm
in the RMF set NL3, and of 2.41 MeV and 0.020 fm in the Gogny force D1S.

To constrain the two parameters $x_0$ and $W_0$ from finite nuclei using the
aforementioned protocol, namely, BE of
$^{208}$Pb and the $1p_{3/2}$--$1p_{1/2}$ level splitting in $^{16}$O
shows some arbitrariness in the sense that
small variations in $W_0$ that imply an appropiate change in $x_0$, do not
modify the
quality of the overall {\it rms} deviations of binding energies and radii. For example,
$x_0$=0.6 and $W_0$=115 MeV give {\it rms} deviations of binding energies and
radii of 1.54 MeV and 0.015 fm, whereas the combination $x_0$=0.2 and $W_0$=116 MeV gives {\it rms} 
deviations of 1.47 MeV and 0.015 fm. It is to be noted here that $t_0$
determined from the BE of $^{40}$Ca is almost insensitive to 
the choice of $x_0$ (because $N=Z$) and $W_0$ (because $^{40}$Ca is spin saturated). 
However, the uncertainty in 
determining $x_0$ 
from finite nuclei has an important impact on the individual contributions to 
the $SE$, $TE$, $TO$ and $SO$ states of $N-N$ interaction in NM. This uncertainty largely manifests in spin properties of NM, as shall be discussed in the 
forthcoming section.

\section{Results and Discussions}

It is evident from the discussions of the foregoing section, on the procedure 
of fixing the parameters of SEI, that for a given value of $\gamma$ a complete 
study of SNM, PNM, ANM and finite nuclei can be performed only by 
assumming standard values of $\rho_0$, $e(\rho_0)$ and 
$E_{s}(\rho_0)$. The empirical values of these three NM properties used by different
models vary over certain ranges as mentioned in section 1. Out of these three
properties, $e(\rho_0)$ has the minimum uncertainty and for all the models its value
lies in the range $-16\pm0.2$ MeV. As regards the other two NM properties, the majority
of the RMF sets have a value of the saturation density $\rho_0$ to the lower side of the
range $0.17\pm0.03$ fm$^{-3}$ and a value of the symmetry energy $E_{s}(\rho_0)$ in the higher
side of the range 33$\pm$5 MeV. In the non-relativistic mean field theories the value of
$\rho_0$ centers around 0.16 fm$^{-3}$ and $E_{s}(\rho_0)$ to the lower side of the above
mentioned range. In both relativistic and non-relativistic microscopic calculations $\rho_0$
is predicted in the higher side of its range. 

In the present work we shall see that $\rho_0$
is strongly correlated with $\gamma$ that determines the stiffness of the EOS in NM. 
In this context we have first examined the pressure-density ($P \sim \rho$) relation 
in SNM for different values of $\gamma$. The pressure in SNM is calculated as
$P(\rho)= \rho^2\frac {de(\rho)}{d\rho}$ from equation (\ref{eq8}), where we have used the values 
$e(\rho_0)$=$-16$ MeV and Fermi kinetic energy $T_{f_0}$=$\frac{\hslash^2
k_{f_0}^2}{2M}$=37 MeV (corresponding to $\rho_0$=0.161 fm$^{-3}$), with
$k_{f_0}$=$(\frac{3{\pi}^2\rho_0}{2})^{1/3}$ being the Fermi momentum at saturation
density. The $P \sim \rho$ curves for $\gamma$=1/6, 1/3, 1/2 and 2/3 
corresponding to incompressibility of NM, $K(\rho_0)$=207, 226, 245 and 263 MeV, respectively, are verified to pass within the experimentally extracted region 
of Ref. \cite{Danielz02}. 
We shall now obtain all the nine parameters of ANM for each EOS of these four 
$\gamma$ by assumming $E_{s}(\rho_0)$=33 MeV together with the values 
$e(\rho_0)$=-16 MeV and $T_{f_0}$=37 MeV used above for SNM. 
Then the study of finite nuclei is performed for each of these four EOSs 
by adopting the procedure for the determination of $t_0$, $x_0$ and $W_0$ outlined in 
the last section. The results for the deviations with respect to experiment in 
the
binding energies, $\delta{E}$, of 161 even-even spherical nuclei and in the charge
radii, $\delta{r_{ch}}$, of 86 even-even spherical nuclei are calculated for 
the four EOSs corresponding to $\gamma$=1/6, 1/3, 1/2 and 2/3. 
The results of $\delta{r_{ch}}$ of the four EOSs presented in Figures 1. 
The {\it rms} deviations from experiment in
the charge radii $\delta{r_{\it rms}}$ in figure 1 for the four EOSs corresponding to
$\gamma$=1/6, 1/3, 1/2 and 2/3 are 0.036 fm$^{-3}$, 0.017 fm$^{-3}$, 0.038 fm$^{-3}$
and 0.057 fm$^{-3}$, respectively. From this figure,  a strong correlation between 
$\gamma$ and $\rho_0$ on the radius of finite nucleus can be seen. It can be concluded that for a given stiffness of 
NM there is a critical value of $\rho_0$ for which the deviations in the charge radii 
of all these nuclei center around zero giving a minimum {\it rms}
value, $\delta{r_{\it rms}}$, of the deviations. 
A relatively softer (stiffer) EOS with $\rho_0$ corresponding to $T_{f_0}$=37 MeV overestimates 
(underestimate) the radii in finite nuclei. It has been verified that this conclusion
does not change on the choice of either the $e(\rho_0)$ or the $E_{s}(\rho_0)$
values. With the four considered EOSs we have made several calculations by varying $e(\rho_0)$ and
$E_{s}(\rho_0)$ and have searched for the optimal {\it rms} results for the BE and radii
of the considered set of spherical nuclei, following the procedure outlined in section 2. 
The results of the minimum {\it rms} deviations in BE, $\delta{E}_{\it {rms}}$, and
in charge radii, $\delta{r}_{\it rms}$, some nuclear matter properties as well as the SEI
parameters $t_0$, $x_0$ and $W_0$ for the four considered EOSs 
%thus obtained 
are given in
Table 2. The remaining parameters of the SEI corresponding to these four EOSs are given in Table 3.
\begin{figure}
\vspace{0.6cm}
\begin{center}
\includegraphics[width=0.98\columnwidth,clip=true]{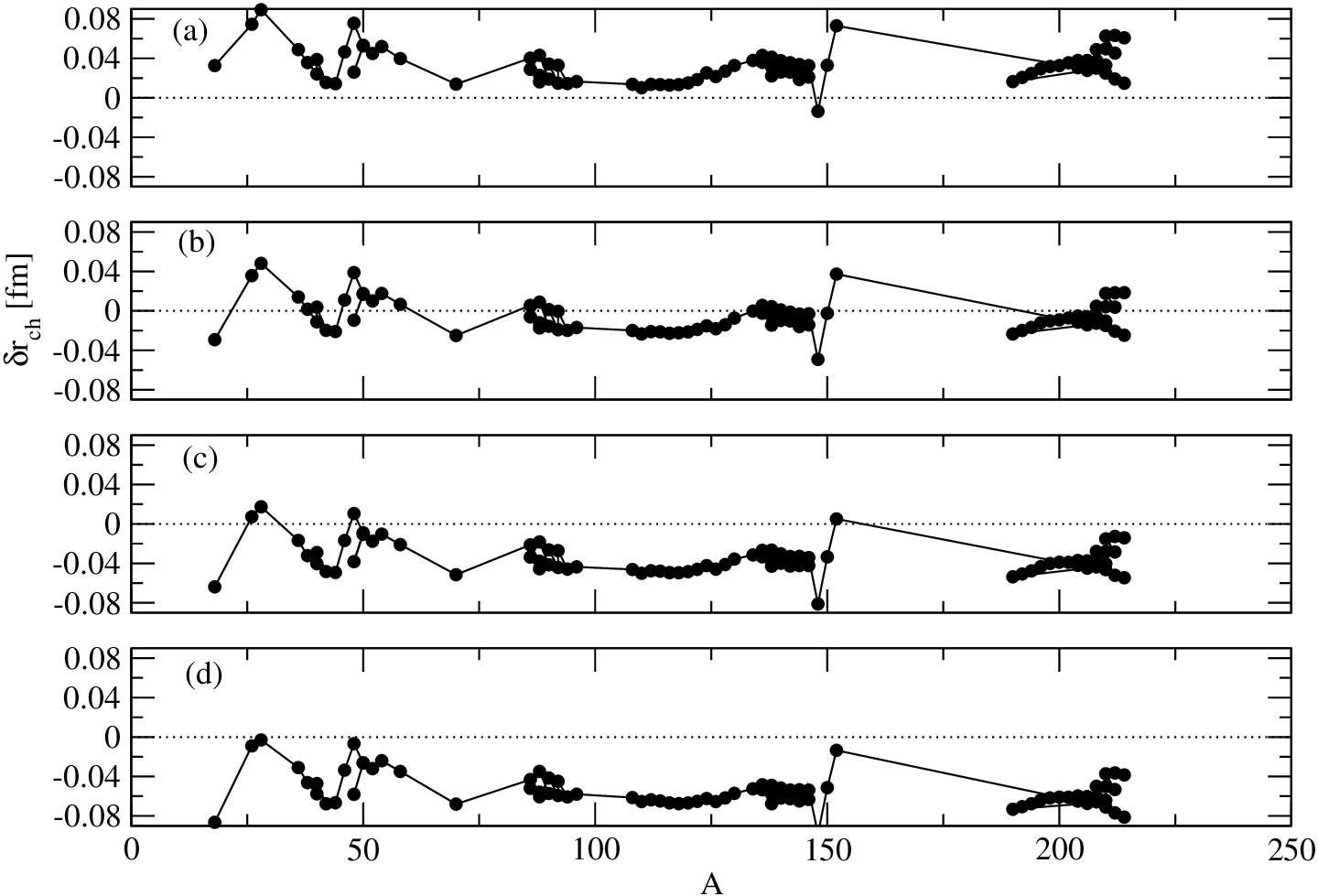}
\caption{Deviation in charge radii, $\delta{r_{ch}}$, for 86 even-even spherical nuclei 
with nucleon number between $A=16$ and $A=224$ for the four EOSs corresponding to the 
same values of $T_{f_0}$=37 MeV and $E_{s}(\rho_0)$ =33 MeV but different 
$\gamma$=1/6, 1/3, 1/2 and 2/3 in (a), (b), (c) and (d), respectively.
}
\end{center}
\label{Fig. 1}
\end{figure}
%%Table 1
\begin{table*} \caption{Values of nuclear matter properties at saturation
(namely, incompressibility, Fermi kinetic energy, density and symmetry energy),
{\it rms} deviations from experiment in energies and charge radii, and the parameters
$t_0$, $x_0$ and $W_0$ for the four EOSs of SEI corresponding to
$\gamma$=1/6, 1/3, 1/2 and 2/3.}
\begin{tabular}{cccccccccc}\hline \hline
$\gamma$ &
$K(\rho_0)$&$T_{f_0}$&$\rho_0$&$E_{s}(\rho_0)$&$\delta{E}_{\it {rms}}$&$\delta{r}_{\it rms}$&$t_0$&$x_0$&$W_0$\\
&MeV&MeV&fm$^{-3}$&MeV&MeV&fm&MeVfm$^3$&&MeV \\ \hline
1/6&207&37.2&0.1623&36&1.4916&0.0187&-575&-0.7&120 \\
1/3&226&36.8&0.1597&35.5&1.3932&0.0167&201&1.1&118 \\
1/2&245&36.4&0.1571&35&1.5402&0.0152&437&0.6&115 \\
2/3&263&36.1&0.1552&35&1.9336&0.0154&540&1.38&112 \\
\hline \hline
\end{tabular}
\end{table*}
%%Table 2
\begin{table*} \caption{Values of the parameters of asymmetric nuclear matter for the four EOSs.}
\begin{tabular}{ccccccccc}\hline \hline
$\gamma$ & $b$&$\alpha$&$\varepsilon_{ex}$&$\varepsilon_{ex}^{l}$&$\varepsilon_{0}$&${\varepsilon_{0}}^{l}$&$\varepsilon_{\gamma}$&$\varepsilon_{\gamma}^{l}$\\
&fm$^3$&fm&MeV&MeV&MeV&MeV&MeV&MeV \\ \hline
1/6&0.2720&0.7568&-96.8427&-64.5618&-215.8954&-131.0415&213.3364&142.7149 \\
1/3&0.4184&0.7582&-95.6480&-63.7653&-112.7493&-67.0819&110.7436&78.7768 \\
1/2&0.5914&0.7597&-94.4614&-62.9743&-78.7832&-45.8788&77.5068&57.7687 \\
2/3&0.7852&0.7609&-93.5766&-62.3844&-61.9929&-33.9536&61.6896&47.0768 \\
\hline \hline
\end{tabular}
\end{table*}

The results of $\rho_0$ in table 2 reveal that for a softer EOS, the central density
is predicted to have a relatively higher value as compared to that of a stiffer EOS. A
similar behaviour is also observed for the symmetry energy. As the incompressibility
of NM changes from 207 MeV to 263 MeV, the saturation density decreases from 0.1623
fm$^{-3}$ to 0.1552 fm$^{-3}$ (corresponding to a decrease of the Fermi kinetic energy
from 37.2 MeV to 36.1 MeV), and the symmetry energy changes from 36 MeV to 35 MeV.
This correlation between $K(\rho_0)$ and $\rho_0$ is also found in earlier
Skyrme II-VI sets \cite{bein75} and conforms to the fact that as the matter
becomes stiffer, the internucleon separation increases. This conclusion is also
substantiated if one examines the values of the incompressibility and the saturation
density of various 
%interaction 
parameter sets of the RMF model that are successfully applied to
finite nuclei calculations (e.g., the popular NL3 set \cite{lala97} has a saturation
density 0.148 fm$^{-3}$ and an incompressibility 271 MeV).
%Figure 4.(a) & (b)
\begin{figure}
\vspace{0.6cm}
\begin{center}
\includegraphics[width=0.98\columnwidth,clip=true]{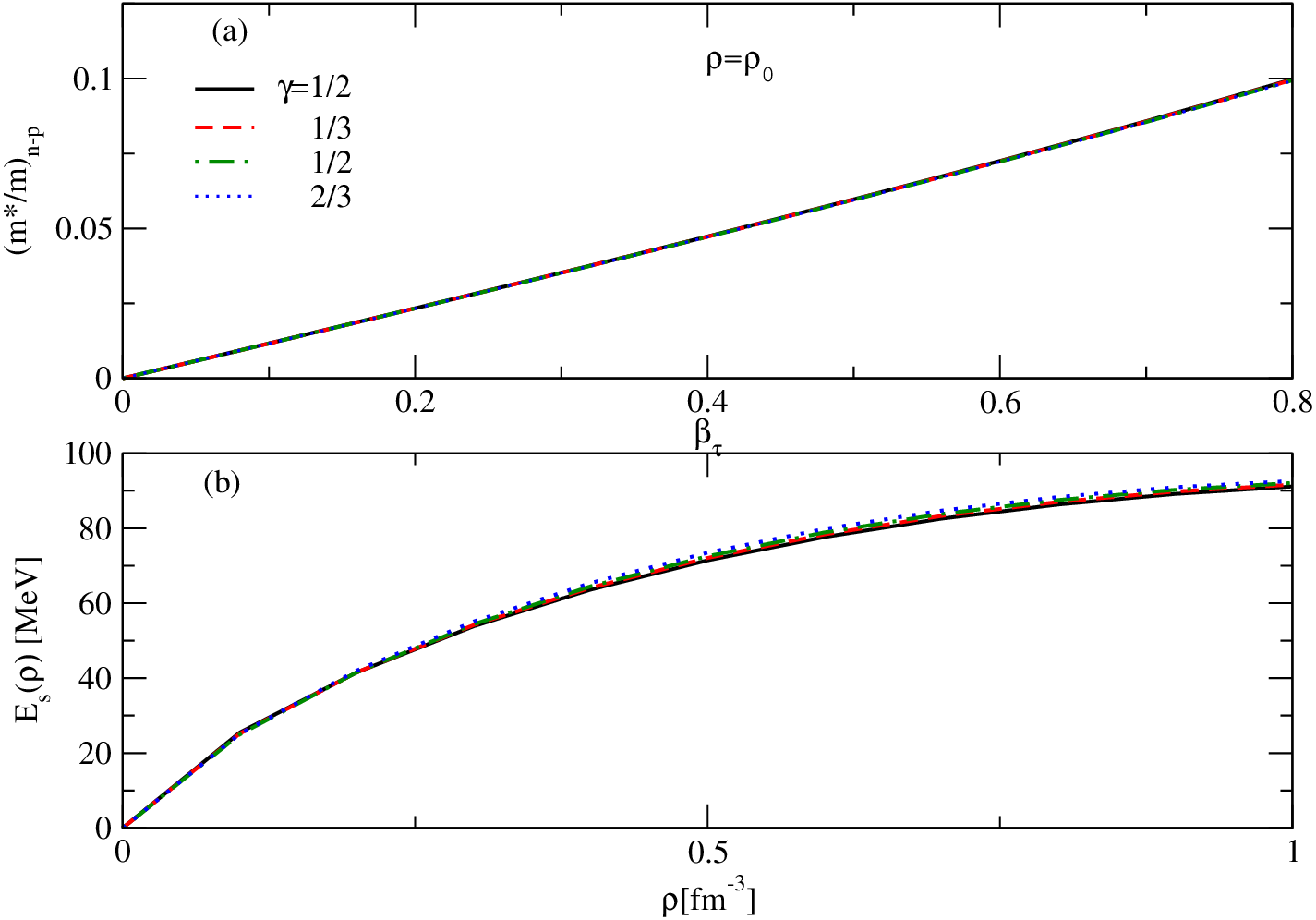}
\caption{(Color online) (a)({\it upper pannel}) The neutron-proton effective mass splitting in
neutron-rich ANM shown as a function of the isospin asymmetry $\beta_{\tau}$
at normal density for the four EOSs of table 2. (b)({\it lower pannel})
The density dependence of the symmetry energy is given for these four EOSs.
}
\end{center}
\label{Fig. 2}
\end{figure}

The variations in the values of $\rho_0$ and $E_{s}(\rho_0)$ with the 
stiffness of NM as ascertained from the study of finite nuclei
have rather small influence on the ANM results. This is 
shown in Figure 2, where the n-p effective mass splitting 
$(m^*/m)_{n-p}$ in ANM at normal density as a function of 
the isospin asymmetry, $\beta_{\tau}$ =$(\rho_n-\rho_p)/(\rho_n+\rho_p)$, 
and the density dependence of symmetry energy are 
given in the upper and lower panels, respectively, for the four sets of EOSs. The
identical results for the n-p effective mass splitting for all four EOSs can be
understood from the almost same values of the exchange strength parameters
$\varepsilon_{ex}$,  $\varepsilon_{ex}^{l}$ and range $\alpha$, given in table 3,
for the EOSs. These parameters determine the momentum dependence of the
mean fields in SNM and PNM. In the process of determination of the exchange
strengths $\varepsilon_{ex}$ and $\varepsilon_{ex}^{l}$ in SNM and PNM, two
particular combinations $(3V_{0}^{TE}-V_{0}^{SO})$ and $(V_{0}^{SE}-3V_{0}^{TO})$
of the strenghts of the finite range part of the interaction in the four basic
states of the N-N interaction are getting fixed (see equations (\ref{eq2.e}) and
(\ref{eq2.f})). These two combinations in terms of the known interaction parameters
are given as,
\begin{eqnarray}
(3V_{0}^{TE}-V_{0}^{SO})=\frac{8\varepsilon_{ex}}{({\rho_0}{\pi^{3/2}}
{\alpha^{3}})}, 
\label{eq21}
\end{eqnarray}
\begin{eqnarray}
(V_{0}^{SE}-3V_{0}^{TO})=\frac{8\varepsilon_{ex}}{3({\rho_0}{\pi^{3/2}}
{\alpha^{3}})}, 
\label{eq22}
\end{eqnarray}
where we have used that $\varepsilon_{ex}^{l}$ = $\frac{2}{3}\varepsilon_{ex}$
and $\varepsilon_{ex}^{ul}$ = $\frac{4}{3}\varepsilon_{ex}$. 

The four EOSs considered here having different stiffness have nearly identical
values of $\varepsilon_{ex}$ (and $\varepsilon_{ex}^{l}$) and, therefore, shall have
a similar momentum dependence of the mean fields in SNM (and PNM). The small differences
are due to the variation in their values of $\rho_0$ which lie within a close range.
Therefore the combinations $(3V_{0}^{TE}-V_{0}^{SO})$ and
$(V_{0}^{SE}-3V_{0}^{TO})$ have similar values for all the four EOSs.

In the course of the determination of the strength parameters
$\varepsilon_{0}$ in SNM and $\varepsilon_{0}^{l}$ 
in PNM, the interaction parameters involved are mutually
adjusted subject to the constraint that $(3V_{0}^{TE}-V_{0}^{SO})$ and
$(V_{0}^{SE}-3V_{0}^{TO})$ retain the values specified by equations (\ref{eq21}) and
(\ref{eq22}). This results into unconstrained variations of the strengths in the
four states of the N-N interaction for the four different EOSs considered. This is
shown in Figure 3 where the individual contributions to the interaction part 
of energy per particle in SNM, $<V>/A$, in the four basic states of $N-N$ 
interaction 
are shown as functions of the Fermi momentum. It can be seen that the contributions
do not follow any definite pattern with respect to their NM properties,
in particular, with increase in stiffness of SNM. The $TO$ state contribution
of the EOS with $\gamma$=1/3 is predicted to be attractive, whereas it is
repulsive for the three other EOSs, being more attractive for $\gamma$=1/6
than in 1/2 case. Similar arbitrary behaviour is observed in $SO$ and $TE$
channels, where, the curves of the EOSs of different stiffness do not follow
any definite trend.
%%Figure 5
\begin{figure*}
%\vspace{0.6cm}
\begin{center}
\includegraphics[width=0.98\columnwidth,clip=true]{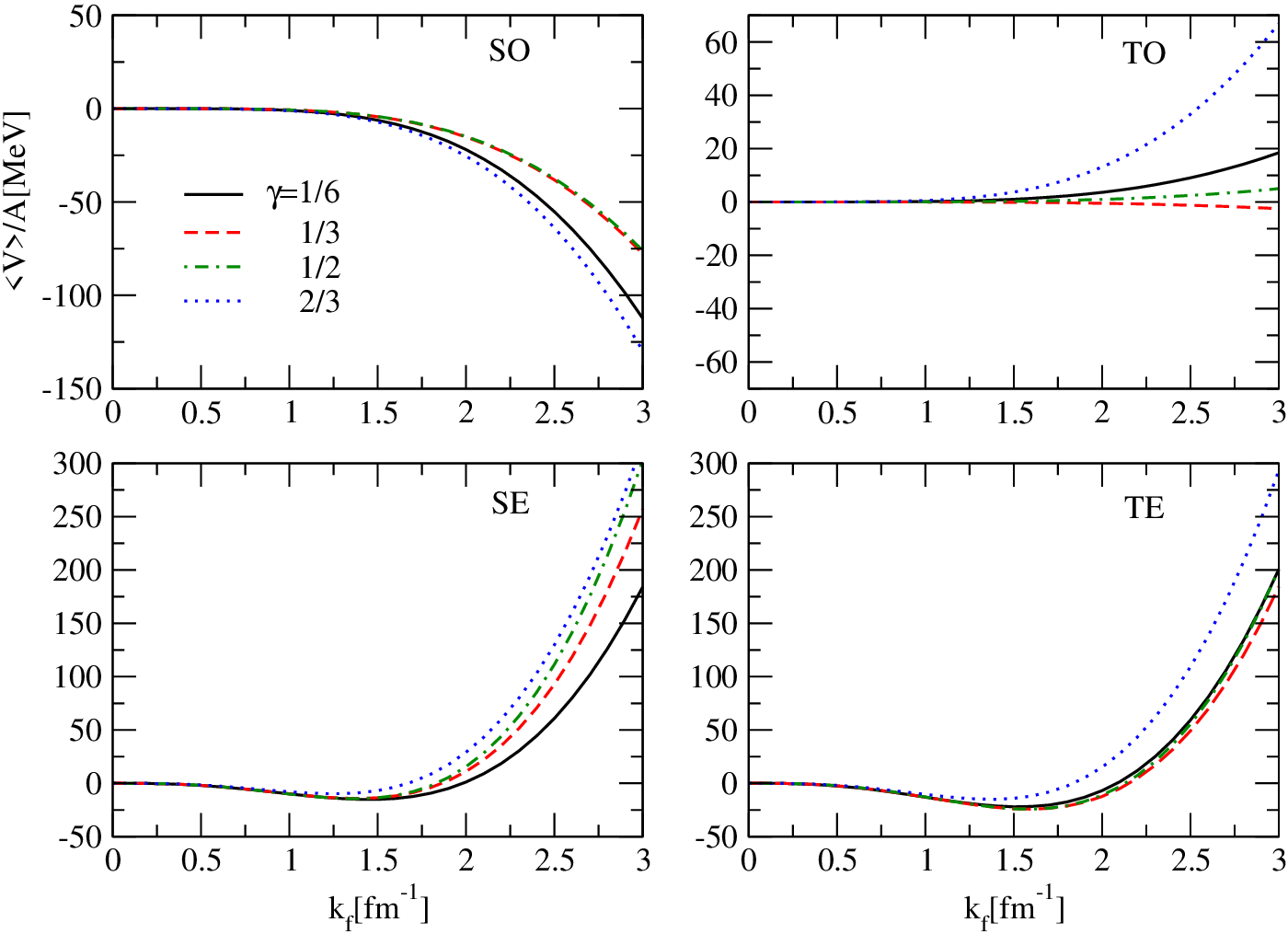}
\caption{(Color online) Contributions to the interaction part of the energy per particle, $<V>/A$,
in SNM coming from the four basic states of the N-N interaction,
${SE}$, ${TE}$, ${TO}$ and ${SO}$, as a function of the Fermi momentum for the four
EOSs corresponding to $\gamma$=1/6, 1/3, 1/2 and 2/3.
}
\end{center}
\label{Fig. 3}
\end{figure*}

%Further we have also verified that 
As mentioned in the previous section, for a given EOS, a small readjustment in
the values of the parameters $x_0$ and $W_0$ (subject to the constraints of
the BE of $^{208}$Pb and the $1p_{3/2}$--$1p_{1/2}$ level splitting in $^{16}$O) 
allows to describe BEs and charge radii of spherical nuclei with similar {\it rms} 
deviations.
%is allowed without appreciably compromising the BEs and radii of finite nuclei.
For example, $x_0$=0.2 and $W_0$=116 MeV could be an alternative set for the
one given in table 2 ($x_0$=0.6 and $W_0$=115 MeV) for $\gamma$=1/2. 
%that predicts the BEs and radii
%of the considered nuclei with {\it rms} deviations 1.4732 MeV and 0.0153 fm respectively
%(in comparison to 1.5402 MeV and 0.0152 fm given in table 1). 
Though these two
EOSs for the same $\gamma$ give identical results in the isospin channel
of ANM (same n-p effective mass splitting and same density dependence of the
symmetry energy), 
%and similar predictions in finite nuclei as well,
the behaviour of their contributions in the four basic channels of the N-N
interaction are found to differ apreciably. This is manifested particularly in the spin channel and the spin symmetry energy $E_{\sigma}(\rho)$ (calculated using the
expression in equation (17) of our earlier work \cite{trr13}).
This is illustrated in Figure 4, where the contribution of the $TO$ state to the interaction energy in SNM
and the spin symmetry energy are plotted as functions of the density in the upper
and lower panels, respectively, for the two sets of values for $x_0$ and $W_0$ corresponding to the EOS of $\gamma$=1/2.
The same situation happens for each of the four EOSs considered in the work. It may be noted that the predictions in the spin channel are crucial in the studies of magnetic
properties of dense NM. Spin polarisation properties in various types of NM
have been studied extensively in theoretical approaches using both microscopic
and
effective models \cite{vid84,kuts94,pandh72,uma97,fan01,vid02,bom02,sam07,sam10,sam11,zuo03,isa04,isa204,isa06,bod07,big10,rio05}, often with contradictory
conclusions. The spin polarizability of NM can have strong effects on
the neutrino mean free path and can impact on the formation mechanism and
cooling scenario of neutron stars. 
%%%%
\begin{figure}
\vspace{0.6cm}
\begin{center}
\includegraphics[width=0.98\columnwidth,clip=true]{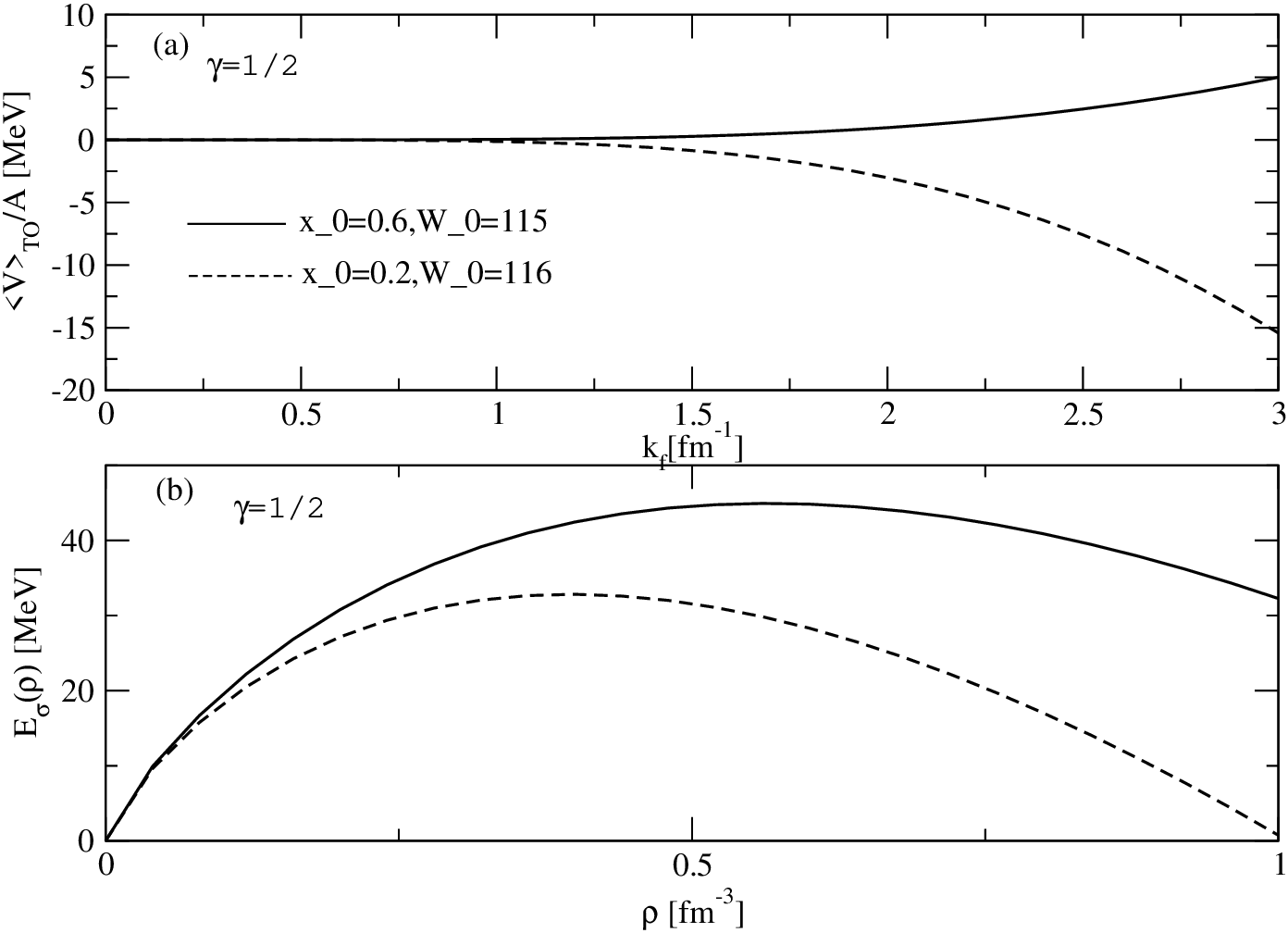}
\caption{(a)({\it {upper panel}}) Contributions to the interaction part of the energy
per particle $<V>^{TO}/A$ in SNM coming from the ${TO}$ state as a function of the
Fermi momentum for the two EOSs having the same $\gamma$=1/2 but differing in their
$x_0$ and $W_0$ values. (b)({\it {lower panel}}) The density dependence of the spin
symmetry energy for the same two EOSs of $\gamma$=1/2.
}
\end{center}
\label{Fig. 4}
\end{figure}

The divergent predictions in the spin channel of the different EOSs for a given $\gamma$ value arises due to the freedom allowed in the determination of the two parameters
$x_0$ and $W_0$ from finite nuclei without compromising much the predictions
of BEs and radii. This results into several possible values
of $x_0$, and hence different sets of values for $V_{0}^{SE}$, $V_{0}^{TE}$,
$V_{0}^{SO}$ and $V_{0}^{TO}$ subject to the constraint that the combinations
$(3V_{0}^{TE}-V_{0}^{SO})$ and $(V_{0}^{SE}-3V_{0}^{TO})$ remain
invariant. A possible way to remove the existing arbitrariness
is to determine the parameter $x_0$ from NM keeping only $W_0$
open for finite nuclei (apart from $t_0$ that is fixed from BE of $^{40}$Ca)
provided the overall predictions in finite nuclei do not worsen.

\subsection{Determination of $x_0$}

The parameter $x_0$ can be expressed as,
\begin{eqnarray}   
x_0=1-{\frac{2}{\rho_0t_0}}(\varepsilon_{0}^{l}-{\varepsilon_{ex}^{l}}+\frac{3}{2}\varepsilon_{ex}^{l,l}),
\label{eq23}
\end{eqnarray}
from equations (\ref{eq2.b}), (\ref{eq2.f}) and (\ref{eq16}). With $t_0$ fixed
from finite nucleus and $\varepsilon_{0}^{l}$ and $\varepsilon_{ex}^{l}$ known from ANM,
we can determine $x_0$ with the knowledge of $\varepsilon_{ex}^{l,l}$. The parameter  $\varepsilon_{ex}^{l,l}$
decides the momentum dependence of the mean field in CPNM and can be
ascertained from the splitting of the parameter  $\varepsilon_{ex}^{l}$
of PNM between spinwise like and unlike pairs in PPNM.
This splitting is subject to the constraint $\varepsilon_{ex}^{l,l}$+$\varepsilon_{ex}^{l,ul}$= $2\varepsilon_{ex}^{l}$. Thus the splitting into the spinwise-like
channel, $\varepsilon_{ex}^{l,l}$, can take any value between 0 and $2\varepsilon_{ex}^{l}$ and
correspondingly $\varepsilon_{ex}^{l,ul}$ is decided.
We have examined the effective mass splitting, $(m^*/m)_{{nu}-{nd}} \equiv
(m^*/m)_{nu}-(m^*/m)_{nd}$,
between spin-up (nu) and spin-down (nd) neutrons in PPNM
at normal density $\rho_0$ for various
possible values of $\varepsilon_{ex}^{l}$, and have compared the results with
the DBHF prediction with the Bonn B potential \cite{sam07}.
The effective masses of $nu$ and $nd$ neutrons in PPNM can be calculated from
the expression,
\begin{eqnarray}   
\bigg [\frac{m^*}{m}(k,\rho,\beta_{\sigma}) \bigg]_{nu,nd}
&=&\bigg [1+{\frac{m}{\hslash^{2}k}}\frac{\partial{u_{nu,nd}
(k,\rho,\beta_{\sigma})}}{\partial{k}} \bigg]^{-1},
\label{effm}
\end{eqnarray}
where $u_{{nu}(nd)}$ is the mean field of $nu$ ($nd$) neutrons in PPNM and $\beta_{\sigma}$ is the spin asymmetry defined as
$\beta_{\sigma}$=$(\rho_{nu}-\rho_{nd})/\rho$,
with $\rho_{nu}$ and $\rho_{nd}$ being the densities of $nu$ and $nd$ neutrons and 
$\rho=\rho_{nu}+\rho_{nd}$, the total density of PPNM. 
The mean fields of $nu$ and $nd$ neutrons in PPNM calculated with the SEI give the results, 
\begin{eqnarray}
u_{nu}(k,\rho,\beta_{\sigma}) &=&
\frac{\varepsilon_{0}^{l,l}}{\rho_0} \rho_{nu}
+ \frac{\varepsilon_{0}^{l,ul}}{\rho_0}\rho_{nd}
+ \frac{\varepsilon_{\gamma}^{l,ul}}{\rho_0^{\gamma+1}}\rho_{nd}
\bigg( 1+\frac{\rho_{nu}}{\rho} \frac{\gamma}{1+b\rho} \bigg)
\left(\frac{\rho}{1+b\rho}\right)^{\gamma}
\nonumber \\
&+&\frac{\varepsilon_{ex}^{l,l}}{\rho_0} \rho_{nu} I(k,k_{nu})
+ \frac{\varepsilon_{ex}^{l,ul}}{\rho_0} \rho_{nd} I(k,k_{nd})
\label{mfnu}
\end{eqnarray}
and
\begin{eqnarray}
u_{nd}(k,\rho,\beta_{\sigma}) &=&
\frac{\varepsilon_{0}^{l,l}}{\rho_0} \rho_{nd}
+ \frac{\varepsilon_{0}^{l,ul}}{\rho_0}\rho_{nu}
+ \frac{\varepsilon_{\gamma}^{l,ul}}{\rho_0^{\gamma+1}}\rho_{nu}
\bigg( 1+\frac{\rho_{nd}}{\rho} \frac{\gamma}{1+b\rho} \bigg)
\left(\frac{\rho}{1+b\rho}\right)^{\gamma}
\nonumber \\
&+&\frac{\varepsilon_{ex}^{l,l}}{\rho_0} \rho_{nd} I(k,k_{nd})
+ \frac{\varepsilon_{ex}^{l,ul}}{\rho_0} \rho_{nu} I(k,k_{nu}) ,
\label{mfnd}
\end{eqnarray}
where $I(k,k_{i})$ ($k_{i}=k_{nu},\,k_{nd}$) is given by equation (\ref{ppnma}) and it has been taken into account that $\varepsilon_{\gamma}^{l,l}=0$ as discussed in Section 2.
Using the SEI mean fields in equation (\ref {effm})
it is found that the $nu$ effective mass, $(m^*/m)_{nu}$, becomes larger than
the $nd$ effective mass, $(m^*/m)_{nd}$, for $\varepsilon_{ex}^{l,l}$ within 0 and
$\varepsilon_{ex}^{l}$, which is the trend observed in the microscopic
DBHF calculation. The difference between the $nu$ and $nd$ effective masses
is maximum for $\varepsilon_{ex}^{l,l}=0$ and this difference decreases
and becomes zero as $\varepsilon_{ex}^{l,l}$ increases in magnitude and
coincides with $\varepsilon_{ex}^{l}$. Beyond this value and up to $\varepsilon_{ex}^{l,l}=2\varepsilon_{ex}^{l}$, the trend reverses and
the $nd$ effective mass becomes larger than the $nu$ effective mass.
Upon comparison with the DBHF result, it is found that for the particular
value of the splitting
$\varepsilon_{ex}^{l,l}$=${\varepsilon_{ex}^{l}}/3$ the results of
$(m^*/m)_{nu-nd}$ for all the four EOSs, corresponding to the four $\gamma$
values, are in close agreement with the microscopic
DBHF prediction \cite{sam07} over a wide range of the spin asymmetry $\beta_{\sigma}$. This is shown in Figure 5, where the calculated results of $(m^*/m)_{nu-nd}$ at saturation density for the value $\varepsilon_{ex}^{l,l}$=${\varepsilon_{ex}^{l}}$/3 along with the DBHF prediction are plotted as a function of the spin asymmetry
$\beta_{\sigma}$.
The momentum dependence of the mean field in spin polarized PNM is fixed once
$\varepsilon_{ex}^{l,l}$ is known. This, in turn, provides us the finite range strength $V_0^{TO}$ in the triplet-odd state which is obtained from equation (\ref{eq16}) using the given value $\varepsilon_{ex}^{l,l}=\varepsilon_{ex}^{l}/3$. The $V_0^{TO}$ strength resulting from this splitting, i.e., $V_0^{TO}=-\varepsilon_{ex}^{l,l}/(\rho_0\pi^{3/2}\alpha^3)$ is found to be
repulsive ($\varepsilon_{ex}^{l}$ is always negative in table 3), and ranges within $V_0^{TO}=54.9-54.6$ MeV as $\gamma$ varies from
1/6 to 2/3. The repulsive character of $V_0^{TO}$ is an essential requirement
for the effective mass in CPNM at normal density to be smaller than 1,
as well as for the stability of CPNM at any higher density.
%Figure 7
\begin{figure}
\vspace{0.6cm}
\begin{center}
\includegraphics[width=0.98\columnwidth,clip=true]{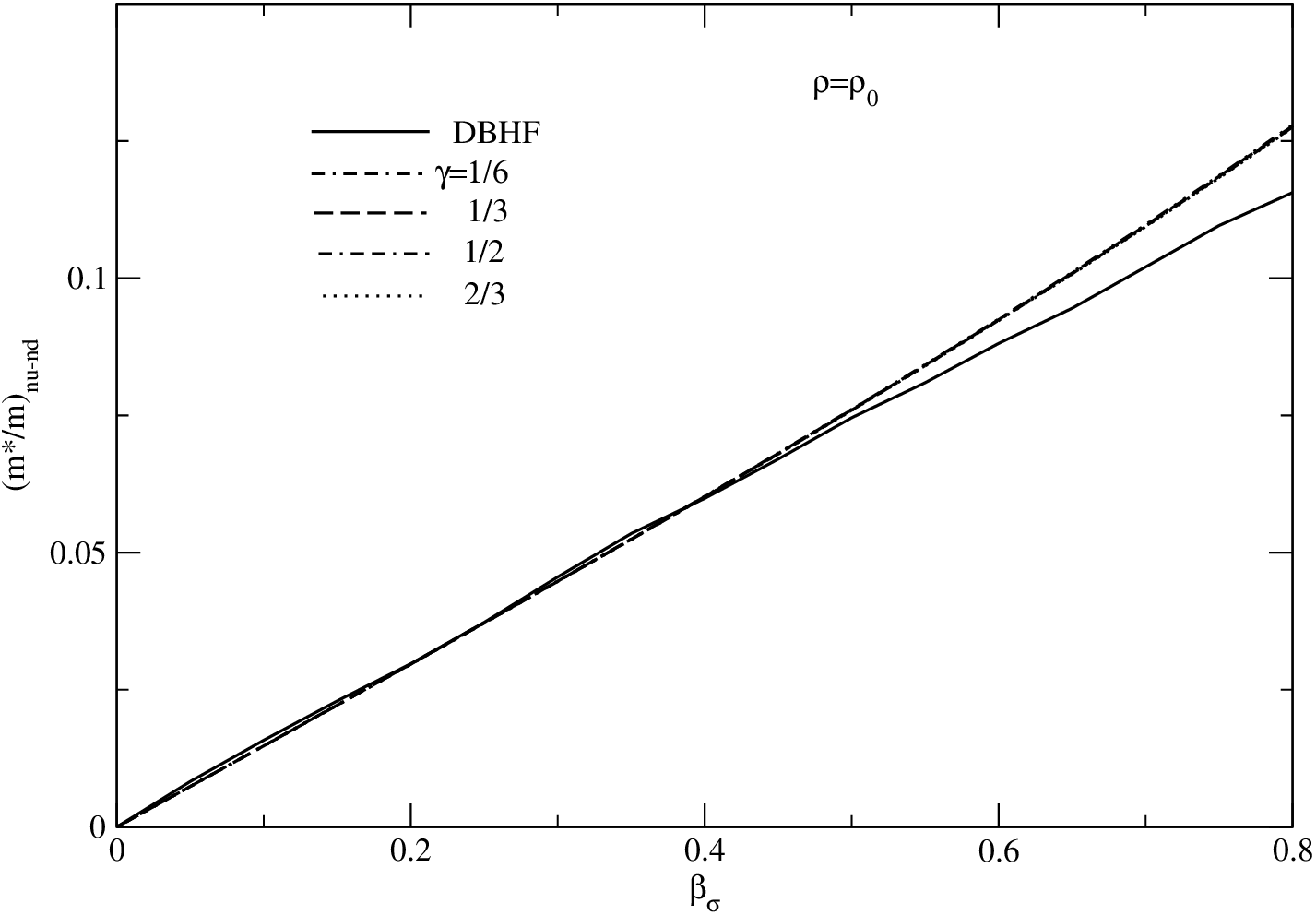}
\caption{(Color online) Effective mass splitting, $(m^*/m)_{nu-nd}$, between spin-up and
spin-down neutrons in polarised PNM as a function of the spin asymmetry $\beta_{\sigma}$=$(\rho_{nu}-\rho_{nd})/(\rho_{nu}+\rho_{nd})$ for the
four EOSs with $\varepsilon_{ex}^{l,l}$=${\varepsilon_{ex}^{l}}/3$. The result of the
microscopic DBHF calculation \cite{sam07} is also given for comparison.
}
\end{center}
\label{Fig. 5}
\end{figure}

The parameter $x_0$, which was determined from the binding energy of $^{208}${Pb} in our earlier fitting procedure \cite{trr13}, now
connects both NM and the finite nucleus. In the present work, once we know
$\varepsilon_{ex}^{l,l}$ from NM and $t_0$ from the finite nucleus,
the parameter $x_0$ is determined from equation (\ref{eq23}).
We fix $t_0$ from the BE of $^{40}${Ca} and the spin-orbit strength $W_0$ from the BE of $^{208}${Pb}. The resulting parameters are reported in table 4.
The $t_0$ values corresponding to different stiffness $\gamma$ of the EOS remain the same as in table 2, whereas $W_0$ changes slightly owing to the change of the $x_0$ parameter, but being free from any arbitrariness. The contributions to the interaction energy in SNM in the four channels resulting from this 
parametrisation are shown in Figure 6. Upon
comparison with the earlier results shown in Figure 3, it can be seen that a
systematic variation with respect to the stiffness of the EOS in SNM is obtained
in the present way of determining the parameters.
The interactions in the $TO$ and $TE$ channels now have a similar behaviour for
all the four EOSs, whereas in the $SO$ and $SE$ channels systematic
variations with respect to the stiffness parameter $\gamma$ are observed.
In both the $SO$ and $SE$ states, the EOS having lower incompressibility
predicts relatively more attraction. In the same figure the results obtained with
different Gogny parameter sets are given for comparison. With the new method of fixing $t_0$ and $x_0$ for SEI, the arbitrariness in the prediction of the spin symmetry energy
$E_{\sigma}(\rho)$ (see lower panel of figure 4) for each EOS also gets removed. The finite range strengths in the
four basic channels $V_{0}^{SE}$, $V_{0}^{TO}$, $V_{0}^{TE}$ and $V_{0}^{SO}$
can be expressed, with the choice $\varepsilon_{ex}^{l,l}$=${\varepsilon_{ex}^{l}}/3$, in terms of $t_0$ as follows:
\begin{eqnarray}
V_{0}^{SE}=\frac{2\varepsilon_{ex}}{({\rho_0}{\pi^{3/2}} 
{\alpha^{3}})},
\label{eq24}
\end{eqnarray}
\begin{eqnarray}
V_{0}^{TO}=-\frac{2\varepsilon_{ex}}{9({\rho_0}{\pi^{3/2}}
{\alpha^{3}})}, 
\label{eq25}
\end{eqnarray}
\begin{eqnarray}
V_{0}^{TE}=\frac{({2\varepsilon_{ex}}+{8\varepsilon_{0}})-6
t_0{\rho_0}}{3({\rho_0}{\pi^{3/2}}{\alpha^{3}})}, 
\label{eq26}
\end{eqnarray}
\begin{eqnarray}
V_{0}^{SO}=\frac{({10\varepsilon_{ex}}+{8\varepsilon_{0}})-6
t_0{\rho_0}}{({\rho_0}{\pi^{3/2}}{\alpha^{3}})}. 
\label{eq27}
\end{eqnarray}
%%Figure 8
\begin{figure*}
%\vspace{0.6cm}
\begin{center}
\includegraphics[width=0.98\columnwidth,clip=true]{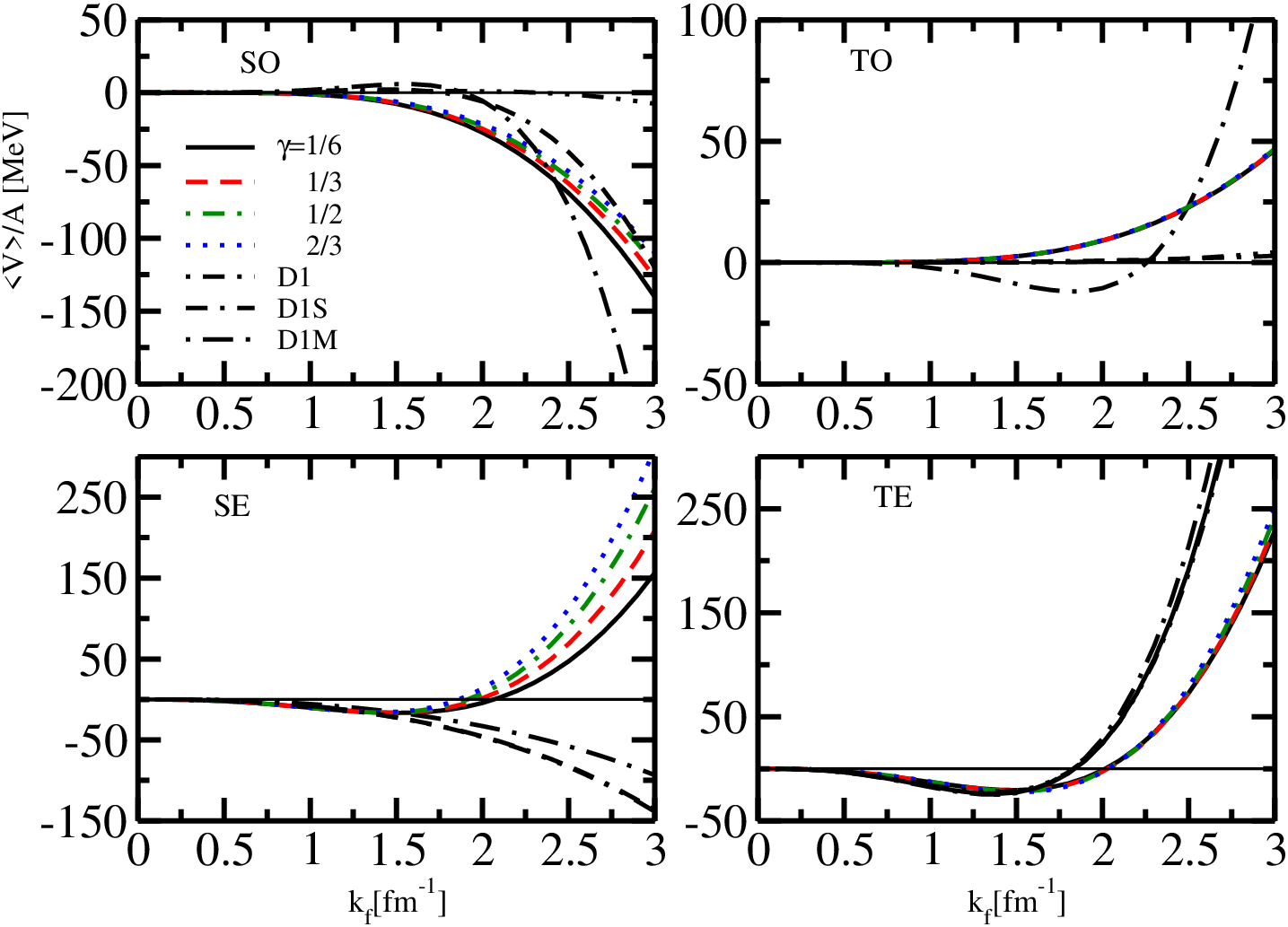}
\caption{(Color online) Contributions from the four basic states of N-N interaction ${SE}$,
${TE}$, ${TO}$ and ${SO}$ to the interaction energy in
SNM as a function of the Fermi momentum for the four EOSs corresponding to
$\gamma$=1/6, 1/3, 1/2 and 2/3 having the parameter $x_0$ determined from the study of
spin polarised PNM and given in table 4.
The results of the Gogny D1, D1S and D1M forces are also shown.
}
\end{center}
\label{Fig. 6}
\end{figure*}
\begin{table*} \caption{Values of the parameters $t_0$, $x_0$ and $W_0$, where $x_0$ is determined from the considerations of both NM and finite nucleus, given in equation (\ref{eq23}),
for the four EOSs along with their {\it rms} deviations in BE of 161 even-even nuclei
and charge radii of 86 even-even nuclei.}
\begin{tabular}{cccccc}\hline \hline
$\gamma$&$t_0$&$x_0$&$W_0$&$\delta{E}_{\it {rms}}$&$\delta{r}_{\it {rms}}$ \\
&MeVfm$^3$&&MeV&MeV&fm \\ \hline
1/6&-575&-1.1161&118&1.6993&0.0189 \\
1/3&201&3.1928&115&1.6754&0.0170 \\
1/2&437&1.4192&112&1.8518&0.0155 \\
2/3&540&1.0659&113.5&1.8297&0.0152 \\
\hline \hline
\end{tabular}
\end{table*}

Once all the parameters of SEI have been fixed, the mean fields
$u_{nu}(k,\rho,\beta_{\sigma})$ and $u_{nd}(k,\rho,\beta_{\sigma})$ of $nu$ and $nd$
neutrons in PPNM can be calculated as a function of the momentum $k$. The mean fields for the SEI set of $\gamma$=1/2 are shown in
Figure 7(a) at density $\rho_0$ and spin asymmetries
$\beta_{\sigma}$=0, 0.1, 0.2 and 0.4. The curves for $nu$ and $nd$ neutrons
lie above and below the unpolarized curve almost symmetrically, with larger separation
between them for larger $\beta_{\sigma}$. The dependence of the $u_{nu}$ and $u_{nd}$ mean fields on the spin asymmetry, calculated at $\rho$=$\rho_0$ and $k$=$k_n$ for the
four EOSs corresponding to $\gamma$=1/6, 1/2, 1/3 and 2/3, is
shown in Figure 7(b) along with the DBHF prediction \cite{sam07}.
In order to have a direct comparison of the results, the curves have been
shifted to the origin by subtracing the respective
$u_{nu(nd)}(k_n,\rho_0,\beta_{\sigma}=0)$ values. The behaviour of the
mean fields of all the four EOSs of SEI is alike but their splittings are smaller
in comparison to the DBHF prediction. The SEI result, with the value $\varepsilon_{ex}^{l,l} = \varepsilon_{ex}^{l}/3$, and the DBHF result have a closely similar momentum dependence in their mean fields in PPNM, as evident from
the effective mass property in
figure 5. Thus, the difference of the results in figure 7(b) can be attributed to
the density-dependent part of the mean fields, that is largely
accounted for by the spin symmetry energy.
In the case of SEI, the splitting of the strength parameter $\varepsilon_{\gamma}^{l}$ of PNM, corresponding to the density-dependent part of the interaction, into the like-spin channel vanishes, i.e., $\varepsilon_{\gamma}^{l,l}=0$, due to the zero range of the density-dependent term. As a consequence, the energy per particle in CPNM cannot have a stiff enough behaviour to ensure that the 
neutrons in polarized state shall have higher value at all densities as 
compared to unpolarized state, the trend obtained in the microscopic BHF and 
DBHF calculations \cite{vid02,sam07}. 
This indicates that the density-dependent part of the interaction in the case of SEI needs to be improved in order to reproduce with better quality the microscopic trend of the density-dependent contribution in spin polarized matter, which shall not be considered in the present work.
%Fig.9a
\begin{figure}
\vspace{0.6cm}
\begin{center}
\includegraphics[width=0.98\columnwidth,clip=true]{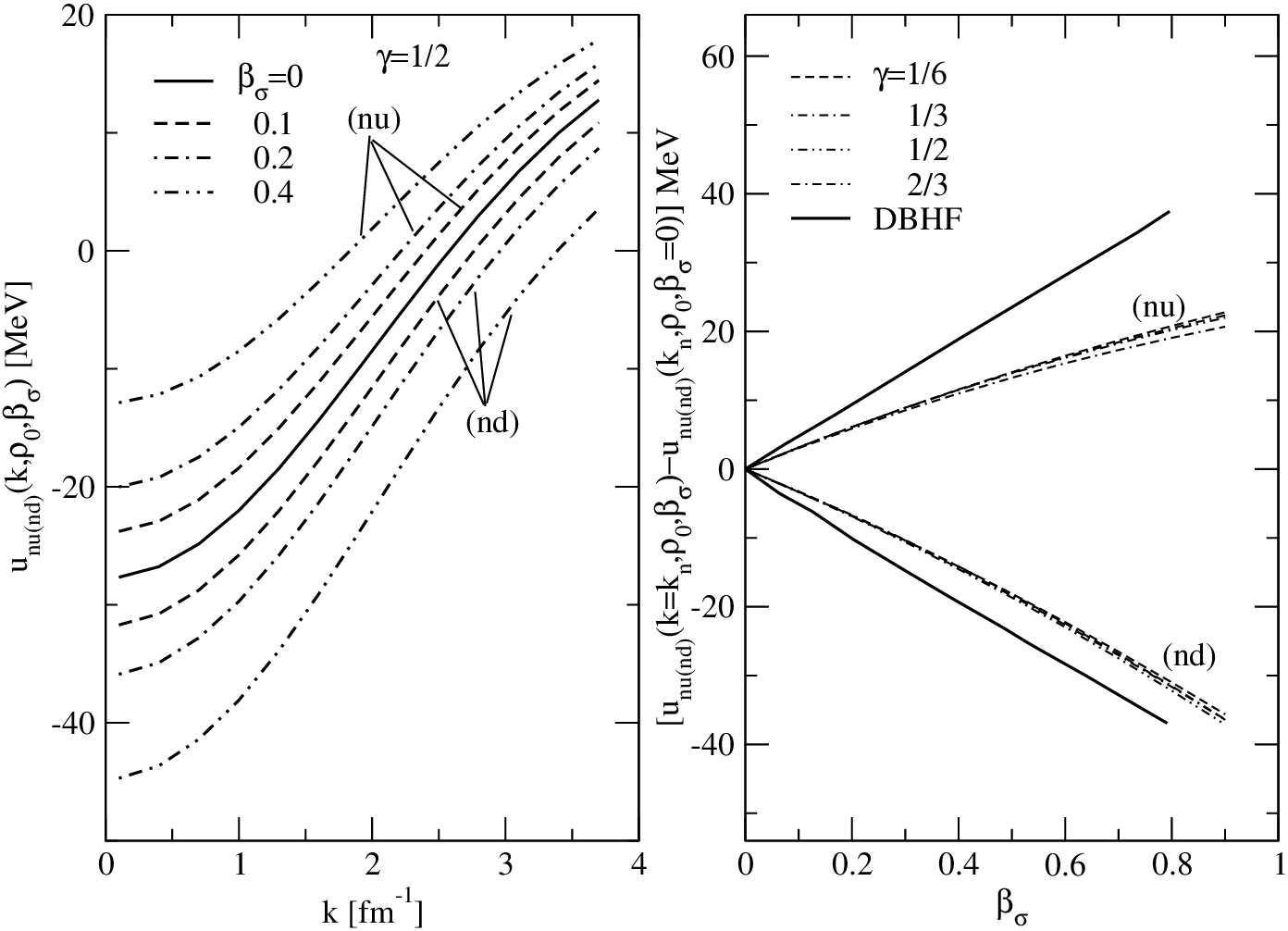}
\caption{(a) Mean fields of $nu$ and $nd$ neutrons in PPNM as a function of momentum $k$ at density $\rho_0$ for the EOS corresponding to $\gamma$=1/2 having values of spin asymmetry $\beta_{\sigma}$=0, 0.1, 0.2 and 0.4. (b) Mean fields of $nu$ and $nd$ neutrons in PPNM at density $\rho_0$ and momentum $k$=$k_n$ as a function of spin asymmetry $\beta_{\sigma}$ for the four EOSs corresponding to $\gamma$=1/6, 1/3, 1/2 and 2/3 alongwith the DBHF results \cite{sam07}.
}
\end{center}
\label{Fig. 7}
\end{figure}

In order to have further insight into the momentum dependence of the mean
fields in PPNM and the effective mass splitting, we note that the sign of the effective mass splitting $(m^*/m)_{{nu}-{nd}}$ is decided by the dimensionless quantity
\begin{eqnarray}
\frac{m}{\hslash^{2}k}\frac{\partial{u^{nu-nd}(k,\rho,\beta_{\sigma})}}{\partial{k}}
\label{dlsefms}
\end{eqnarray}
where $u^{nu-nd}(k,\rho,\beta_{\sigma})$ is the difference
between the $nu$ and $nd$ mean fields given by equations (\ref{mfnu}) and (\ref{mfnd}). If equation (\ref{dlsefms}) is negative (positive) the $nu$ ($nd$) effective mass is larger than the $nd$ ($nu$) effective mass, as can be easily deduced from equation (\ref{effm}).
In the limit $\beta_{\sigma} \to 0$, equation (\ref{dlsefms}) reads as
\begin{eqnarray}
\frac{m}{\hslash^{2}k}\frac{\partial{u^{nu-nd}(k,\rho,\beta_{\sigma})}}
{\partial{k}}\bigg\vert_{k=k_{n}} = -\beta_{\sigma}(\varepsilon_{ex}^{l,l}
-\varepsilon_{ex}^{l,ul})\frac{\rho\Lambda^{2}}{4\rho_{0}{k_n^4}}
\frac{m}{\hslash^{2}}
% \times \nonumber \\
% &&
\bigg[1-\left(1+\frac{4{k_n^2}}{\Lambda^2}\right)e^{-\frac{4{k_n^2}}{\Lambda^2}} \bigg]
\nonumber \\
\label{effm1}
\end{eqnarray}
This equation can predict the
nature of the effective mass splitting in spin polarized PNM
by inspecting the value of ($\varepsilon_{ex}^{l,l}-\varepsilon_{ex}^{l,ul}$).
Due to the fact that the square bracketed factor in equation (\ref{effm1}) is positive and the exchange strength parameters are attractive,
the $nu$ ($nd$) effective mass will lie above the $nd$ ($nu$) effective mass
if ($\varepsilon_{ex}^{l,l}-\varepsilon_{ex}^{l,ul}$) is positive (negative).
In terms of the interaction parameters, we have  ($\varepsilon_{ex}^{l,l}-\varepsilon_{ex}^{l,ul}$)=$-\frac{8\rho_0\pi^{3/2}}{\Lambda^3}(W-H)$, and hence the sign of (\ref{effm1}) can be predicted in terms of the value of $W-H$.
Therefore, to have a $nu$ effective mass larger than the $nd$ effective mass requires that $W-H$ be negative.
In the case of the Gogny interaction, the short-range term being dominant, the behaviour of the $nu$ and $nd$ effective masses can be predicted
from the value of $W_{1}-H_{1}$ (i.e., the quantity $W-H$ in the short-range term of the Gogny forces). In the Gogny D1, D1S, D1N and D1M parameter sets, $W_{1}-H_{1}$ assumes positive values and it has been verified that in these sets the $nd$ neutron effective mass in PPNM lies above the $nu$ effective mass.

\subsection{Spin polarized SNM}

We shall study the two extreme cases of spin polarisation, CSNM and CASNM, 
in SNM. The energy per particle in CSNM and CASNM using the 
SEI is calculated from equations (\ref{eq9}) and (\ref{eq10}) and is shown 
in the Figures 8.(a) and (b), respectively, for the four EOSs corresponding 
to $\gamma$=1/6, 1/3, 1/2 and 2/3 together with the results of the SNM. 
The results in the figures are shown upto 10 times the normal NM density 
$\rho_0$ as the central densities of the maximum mass neutron stars 
obtained in this range of NM incompressibility are found, in an earlier 
work \cite{trr07}, to entend upto 8 to 10 times the normal density. 
Like the isospin symmetry energy in ANM, 
the spin symmetric energy in these two 
types of spin polarized SNM can be expressed as the differences of 
energy per particle in SNM from that of in CSNM (CASNM),
\begin{eqnarray}
E_{\sigma}^{i}(\rho)=e_{pol}^{i}(\rho)-e(\rho), 
\label{spin}
\end{eqnarray}
with $i$=S(AS). The spin symmetry energy thus calculated for these two 
cases of polarisations in SNM are shown in Figures 9.(a) and (b) for the four 
EOSs of SEI alongwith the results of microscopic and effective models. 
The spin symmetry energy for SEI calculated from equation (17) of
Ref.\cite{trr13}, derived under the Taylor series expansion of the energy
per particle in spin asymmetric ANM, is also shown in figure 9.(a) (curves with 
crosses) and is seen to compare well with the results of CSNM.

\begin{figure}
\vspace{0.6cm}
\begin{center}
\includegraphics[width=0.98\columnwidth,clip=true]{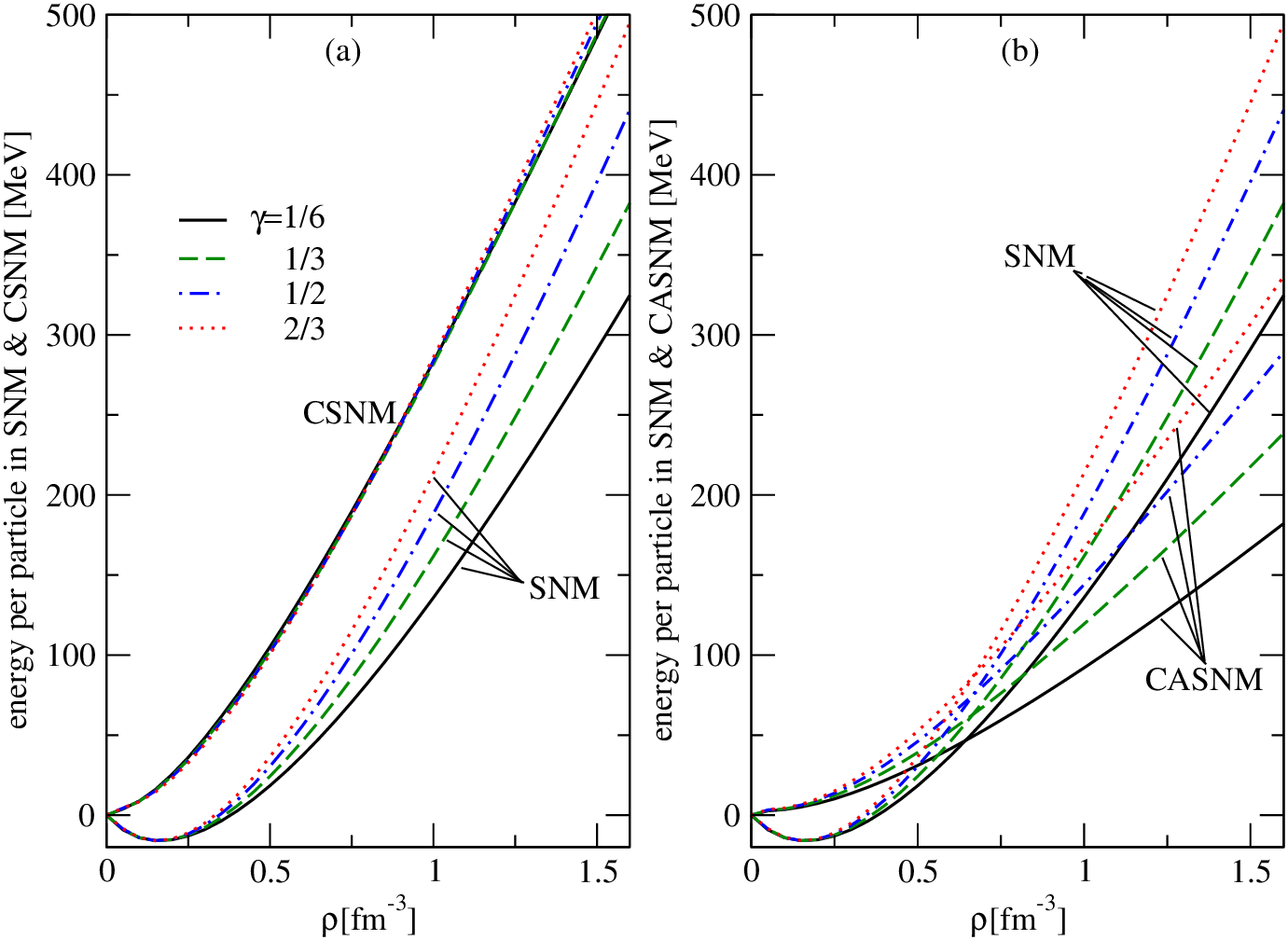}
\caption{(Color online) (a) Energy per particle in CSNM in equation (\ref {eq9}) is shown as functions of density $\rho$ for the four EOSs of SEI corresponding to $\gamma$=1/6, 1/3, 1/2 and 2/3, with the sets of parameters determined for the values of $x_0$ in table 4. (b) Same as (a) but for CASNM in equation (\ref {eq10}).
}
\end{center}
\label{Fig. 8}
\end{figure}
%
%%Figure 10
\begin{figure}
\vspace{0.6cm}
\begin{center}
\includegraphics[width=0.98\columnwidth,clip=true]{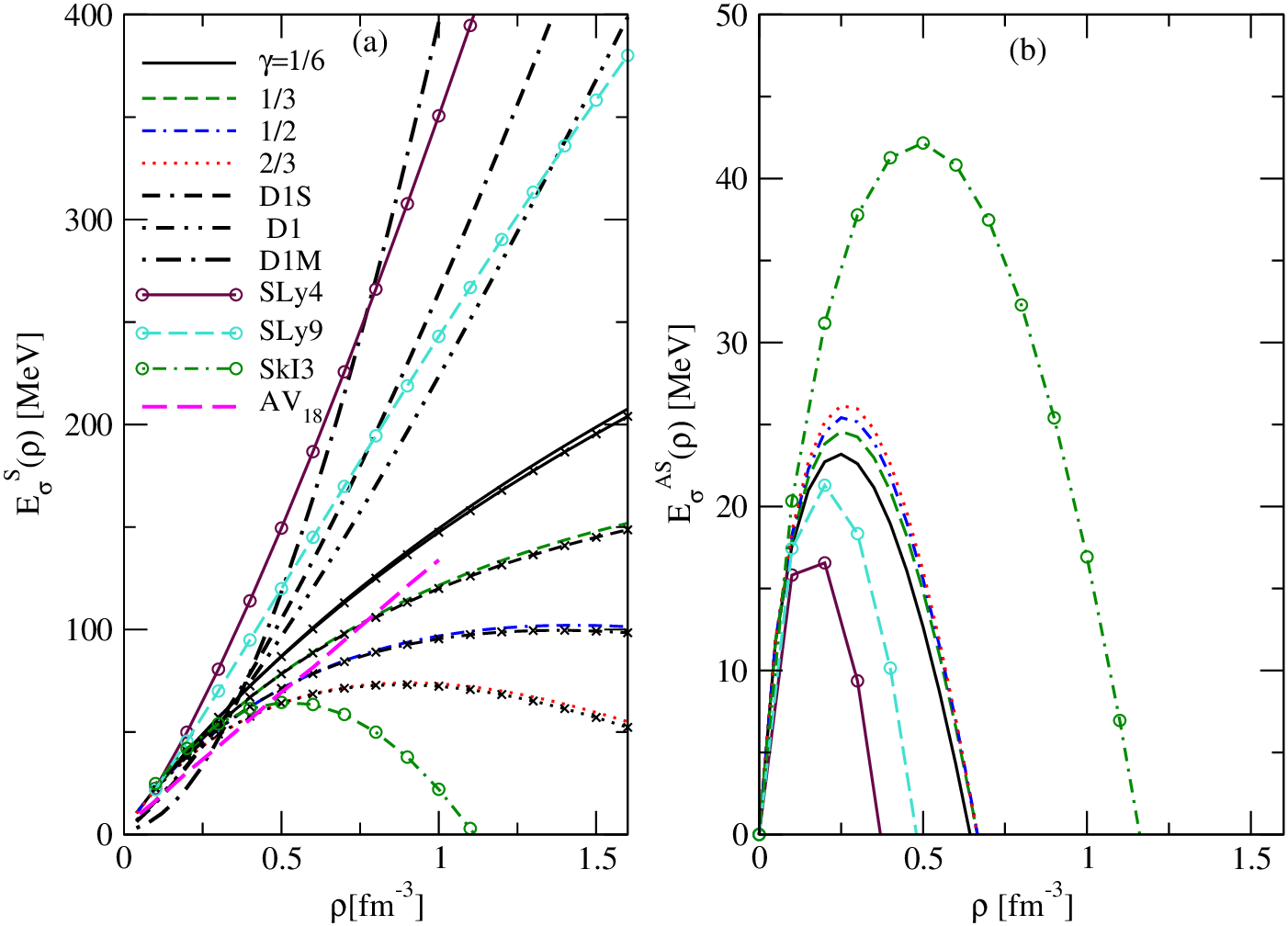}
\caption{(Color online) (a) Spin symmetry energy, $E_{\sigma}^{S}(\rho)$, calculated from equation (\ref{spin}) for FM spin polarised SNM as a function of density for the four EOSs
corresponding to $\gamma$=1/6, 1/3, 1/2 and 2/3 with the sets of parameters determined for values of $x_0$ in table 4. The corresponding results
calculated from the equation (17) of ref.\cite{trr13} are shown
by curves with crosses. The results of the microscopic LOCV calculation with
$AV_{18}$ interaction \cite{bod07} and of the Gogny D1, D1S and D1M \cite{gogny80,blaizot95,goriely09}; Skyrme SLy4, SLy9 and SkI3 force sets \cite{rein95,chab97} are also shown. (b) Same as (a) but for the AFM spin polarised SNM.
}
\end{center}
\label{Fig. 9}
\end{figure}

It can seen from figure 8.(a) and 9.(a) that the SEI force sets do not 
predict FM spin ordering in SNM in agreement with the predictions of all 
microscopic models \cite{bod07,sam11} and some of the effective models. 
However, AFM spin ordering in SNM is predicted by the SEI sets as in case of 
Gogny D1S force (cf. Ref. \cite{isa04}), which is contrary to the 
microscopic predictions. The SEI sets allow the antiferromagnetic transition at 
a density about four times the saturation density in SNM. 
The results for the three considered Skyrme sets, namely SLy4, SLy9 
ans SkI3 \cite {rein95,chab97} show divergent behaviour. For the SLy4 and SLy9 forces, 
a FM transition is not realized at any density, whereas AFM ordering
of the neutron and proton spins in SNM occurs at 
relatively high densities. In the case of SkI3
both FM and AFM spin ordering are possible, with the former being realized at a relatively 
smaller density than the latter.
The Skyrme results are calculated from equation (\ref{spin}), where the expressions
of the energy per particle in CSNM and CASNM are, 
\begin{eqnarray}
e_{pol}^{Sky-S}(\rho) &=& \frac{3\hslash^2{k_f^{pol}}^2}{10M} \nonumber
+\frac{t_{0}}{4}(1+x_0)\rho+\frac{t_{3}}{24}(1+x_3)\rho^{\gamma+1} \\\nonumber
&& +{\bigg[\frac{t_1(1+x_1)+3t_2(1+x_2)}{8} \bigg]}\left(\frac{3}{5}{k_f^{pol}}^2\rho\right) \\
\label{sky}
\end{eqnarray}
and
\begin{eqnarray}
e_{pol}^{Sky-AS}(\rho) &=& \frac{3\hslash^2{k_f^{pol}}^2}{10M} \nonumber
+\frac{t_{0}}{4}\rho+\frac{t_{3}}{24}\rho^{\gamma+1} \\\nonumber
&&+{\bigg[\frac{t_1+t_2(3+2x_2)}{8} \bigg]}\left(\frac{3}{5}{k_f^{pol}}^{2}\rho\right) \\
\label{sky1}
\end{eqnarray}
respectively.

For the sake of further insight, we express
the energy per particle of CSNM and CASNM given in equations (\ref{eq9}) and
(\ref{eq10}) in terms of the finite range strength parameter in the triplet-odd
state $V_0^{TO}$ together with the other known parameters, as given by,
\begin{eqnarray}
e_{pol}^{S}(\rho) &=& \bigg[\frac{3\hslash^2{k_f^{pol}}^2}{10M}
+\frac{\varepsilon_{\gamma}^{ls}}{2\rho_0^{\gamma+1}} 
\rho\left(\frac{\rho({\bf R})} {1+b\rho({\bf R})}\right)^{\gamma} \nonumber \\
&&+ \frac{\rho}{2\rho_0}\bigg(\frac{\rho_0t_0}{2}-\varepsilon_{0}^{l}
+\frac{2}{3}(\varepsilon_{0}+\varepsilon_{ex}) \bigg) \nonumber \\
&&+ \left(\frac{2}{3}\varepsilon_0-\frac{\rho_0t_0}{2}\right)J(k_{f}^{pol}) \bigg] \nonumber \\
&&+ V_0^{TO} \bigg[(\frac{3}{4}\pi^{3/2}{\alpha}^3\rho)\left(1-J(k_{f}^{pol})\right) \bigg] ,
\label{csnm1}
\end{eqnarray}
and
\begin{eqnarray}
e_{pol}^{AS}(\rho) &=& \bigg[\frac{3\hslash^2{k_f^{pol}}^2}{10M}
+\frac{\varepsilon_{\gamma}^{las}}{2\rho_0^{\gamma+1}}
\rho\left(\frac{\rho({\bf R})} {1+b\rho({\bf R})}\right)^{\gamma} \nonumber \\
&&+ \frac{\rho}{2\rho_0}\bigg(-\frac{\rho_0t_0}{2}+\frac{4}{3}\varepsilon_{0}
-\frac{2}{3}\varepsilon_{ex} \bigg)  \nonumber \\
&&+ \left(-\frac{2}{3}\varepsilon_0+\frac{\rho_0t_0}{2}+\frac{4}{3}\varepsilon_{ex}\right)J(k_{f}^{pol}) \bigg] \nonumber \\
&&+ V_0^{TO} \bigg[-(\frac{1}{4}\pi^{3/2}{\alpha}^3\rho)\left(1-J(k_{f}^{pol})\right) \bigg] ,
\label{csnm2}
\end{eqnarray}
respectively, where $J(k_{f}^{pol})$ can be obtained from equation
(\ref{eq10a}) for $k_i$=$k_{f}^{pol}$. 
Now, we can write them as
\begin{eqnarray}
 e_{pol}^{S(AS)}(\rho) &=& A_{s(as)}(\rho) + V_0^{TO}B_{s(as)}(\rho)
\end{eqnarray}
where $A_{s(as)}(\rho)$ and $B_{s(as)}(\rho)$ represent the first and second
square bracketed terms in equation (\ref{csnm1}) (equation (\ref{csnm2})), 
repectively. 
The quantities $A_{s(as)}$ and $B_{s(as)}$ are shown in
Figures 10.(a) and (b), respectively, as a function of density for both CSNM and CASNM.
The results in these two figures show that the energy
per particle in CSNM, $e_{pol}^{S}(\rho)$, shall be an increasing
function of density for repulsive $V_0^{TO}$. The procedure adopted in the
determination
of $x_0$, discussed in the foregoing subsection, also predicts a repulsive
value of $V_0^{TO}$ and hence an increasing trend of the spin symmetry
energy is obtained for the EOSs of SEI, exhibiting stability against
FM spin ordering in SNM. 
The results of $A_s(\rho)$ for the four EOSs obtained under the present
procedure of determination of the parameters are almost identical. Hence,
the CSNM energy per particle $e_{pol}^{S}(\rho)$ shows little dependence on the stiffness of SNM, as can be seen from the curves for the four EOSs of CSNM in
figures 8.(a) and 10.(a). Due to the fact that $e(\rho)$ is stiffer for higher $\gamma$ values, the spin symmetry energy $E_{\sigma}(\rho)$ shall exhibit a softer behaviour and this can be seen from figure 9(a). On the other hand, the energy per particle in CASNM, $e_{pol}^{AS}(\rho)$, has a dependence on the stiffness parameter $\gamma$ of the EOS, but this dependence is not stronger than its counterpart in SNM as one can realize from figure 8.(b). From figure 10.(a) it can be observed that $A_{as}(\rho)$ has a stiffer behaviour for higher $\gamma$ values but it remains below its counterpart curve $A_{s}(\rho)$ of CSNM at all densities. Moreover, the contribution of the $B_{as}(\rho)$ term is negative for a repulsive $V_0^{TO}$ as can be seen
from equation (\ref{csnm2}) and figure 10.(b). This makes the energy per particle
$e_{pol}^{AS}(\rho)$ in CASNM softer than its counterpart $e_{pol}^{S}(\rho)$ in SNM,
and AFM spin ordering is predicted at a critical density close to 0.65 fm$^{-3}$
for all the four EOSs of SEI, as found in figure 9.(b).
It may be pointed out that the AFM spin polarized SNM can
be stable against the AFM transition for an attractive $V_0^{TO}$.
But in that case, the FM spin polarized SNM shall be realized.
It may be mentioned that FM and AFM spin ordering in different types of NM is still under debate and predictions of different model calculations are often 
contradictory (cf. Ref\cite{big10}).

%Figure 11
\begin{figure}
\vspace{0.6cm}
\begin{center}
\includegraphics[width=0.98\columnwidth,clip=true]{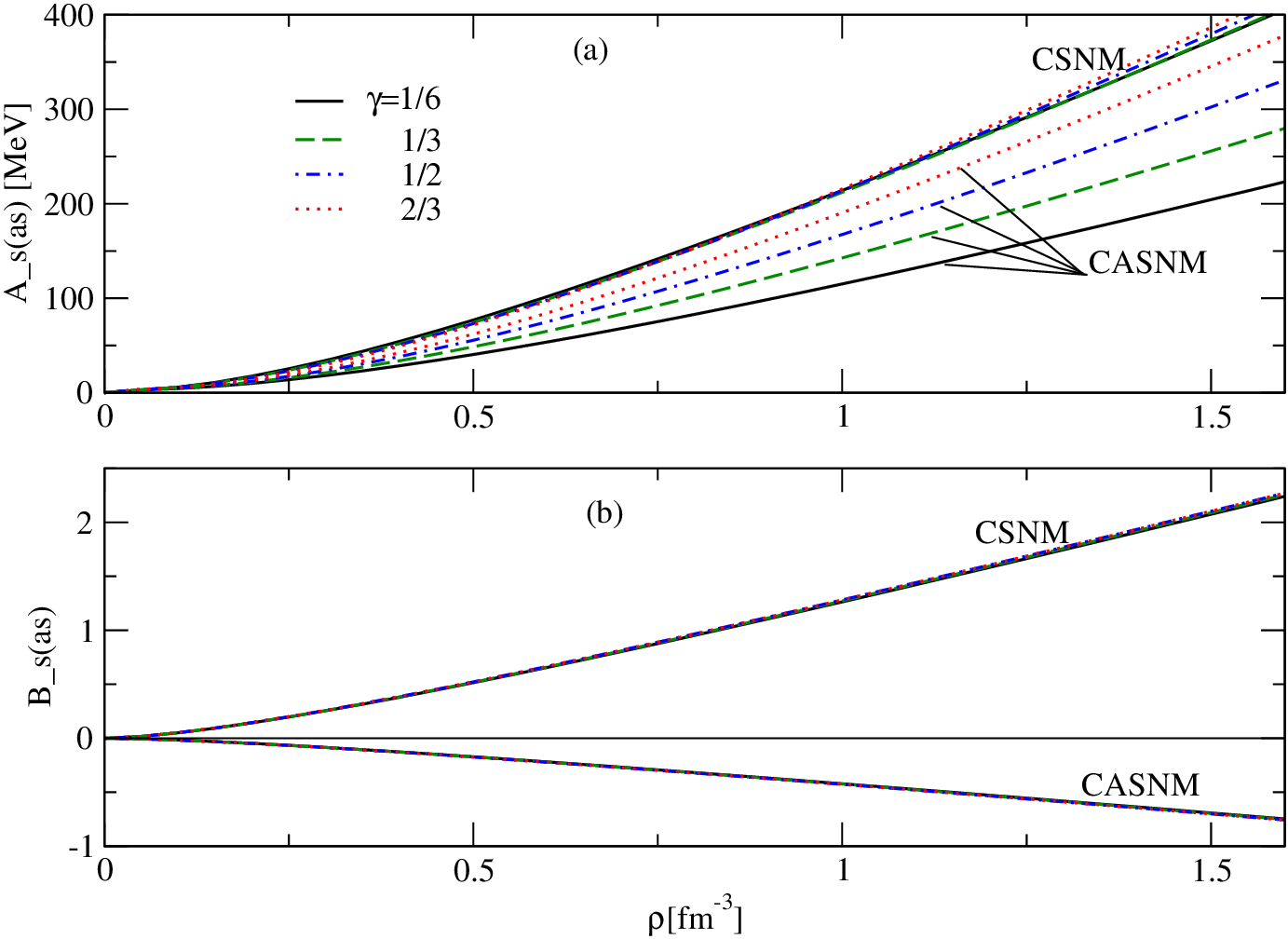}
\caption{(Color online) ({\it (a) upper panel}) The function $A_{s(as)}(\rho)$ defined in equation (\ref{csnm1}) (equation (\ref{csnm2})) for CSNM (CASNM) is shown as function of density $\rho$ for the four sets of EOSs corresponding to $\gamma$=1/6, 1/3, 1/2 and 2/3 with the sets of parameters determined for values of $x_0$ in table 4. ((b)lower panel) The 
function $B_{s(as)}(\rho)$ defined in equation (\ref{csnm1}) (equation (\ref{csnm2})) for 
CSNM (CASNM) as function of density $\rho$ for the same four EOSs as in (a). 
}
\end{center}
\label{Fig. 10}
\end{figure}
%
%
%
%
%%%%%%%%%%%%%%%%%%%%%%%%%%%%%%%%%%%%%%%%%%%%%%%%%%%%%%%%%%%%%%%%%%%%%%%%%%%%
%Figure 12
\begin{figure}
\vspace{0.6cm}
\begin{center}
\includegraphics[width=0.98\columnwidth,clip=true]{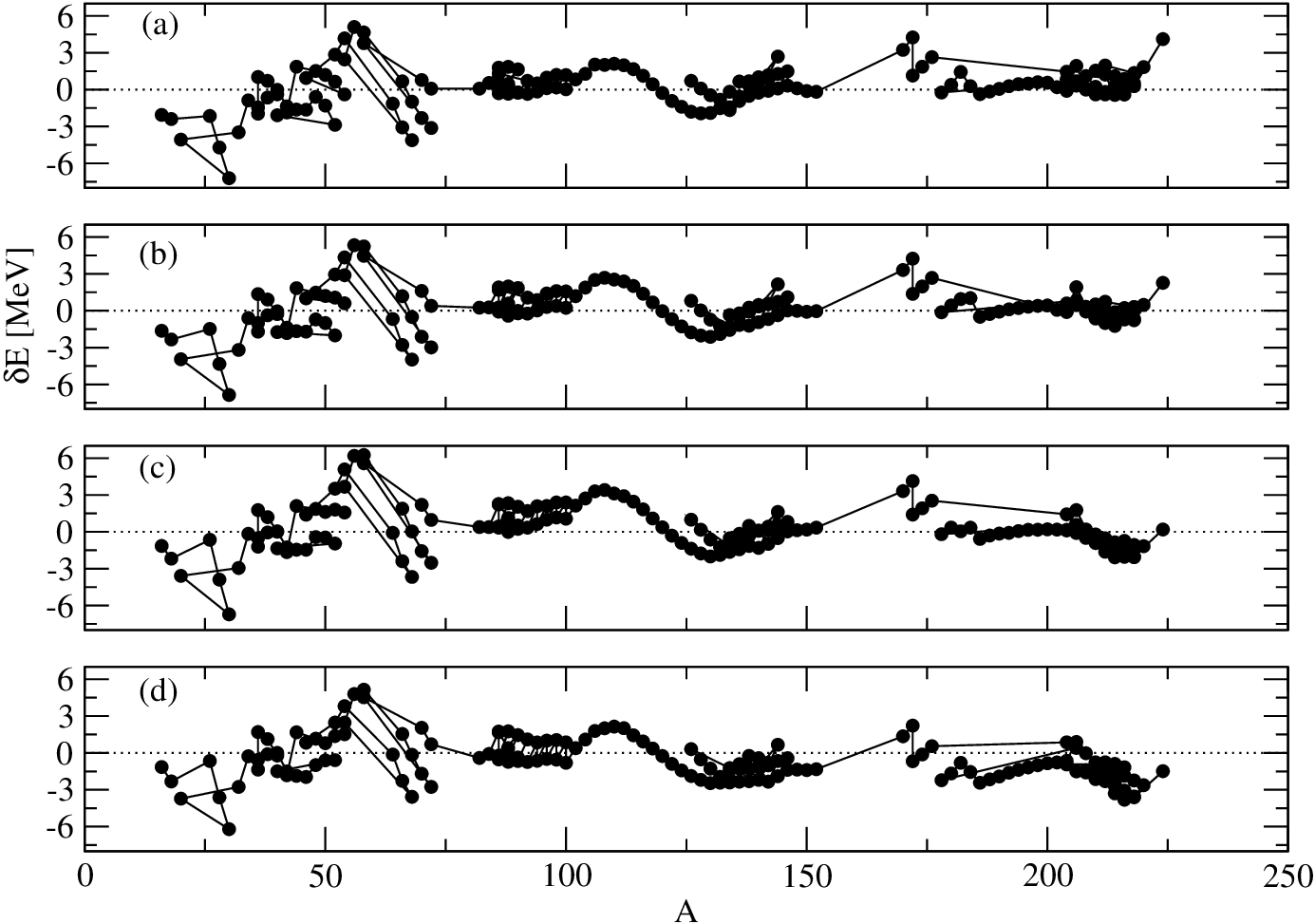}
\caption{Deviation in energy for 161 even-even spherical nuclei with
nucleon number between $A=16$ and $A=224$ for the EOSs corresponding to
$\gamma$=1/6, 1/3, 1/2 and 2/3 with sets of parameters for $x_0$ in table 4 
and are shown in (a), (b), (c) and (d), respectively.
}
\end{center}
\label{Fig. 11}
\end{figure}
%%Figure 13
\begin{figure}
\vspace{0.6cm}
\begin{center}
\includegraphics[width=0.98\columnwidth,clip=true]{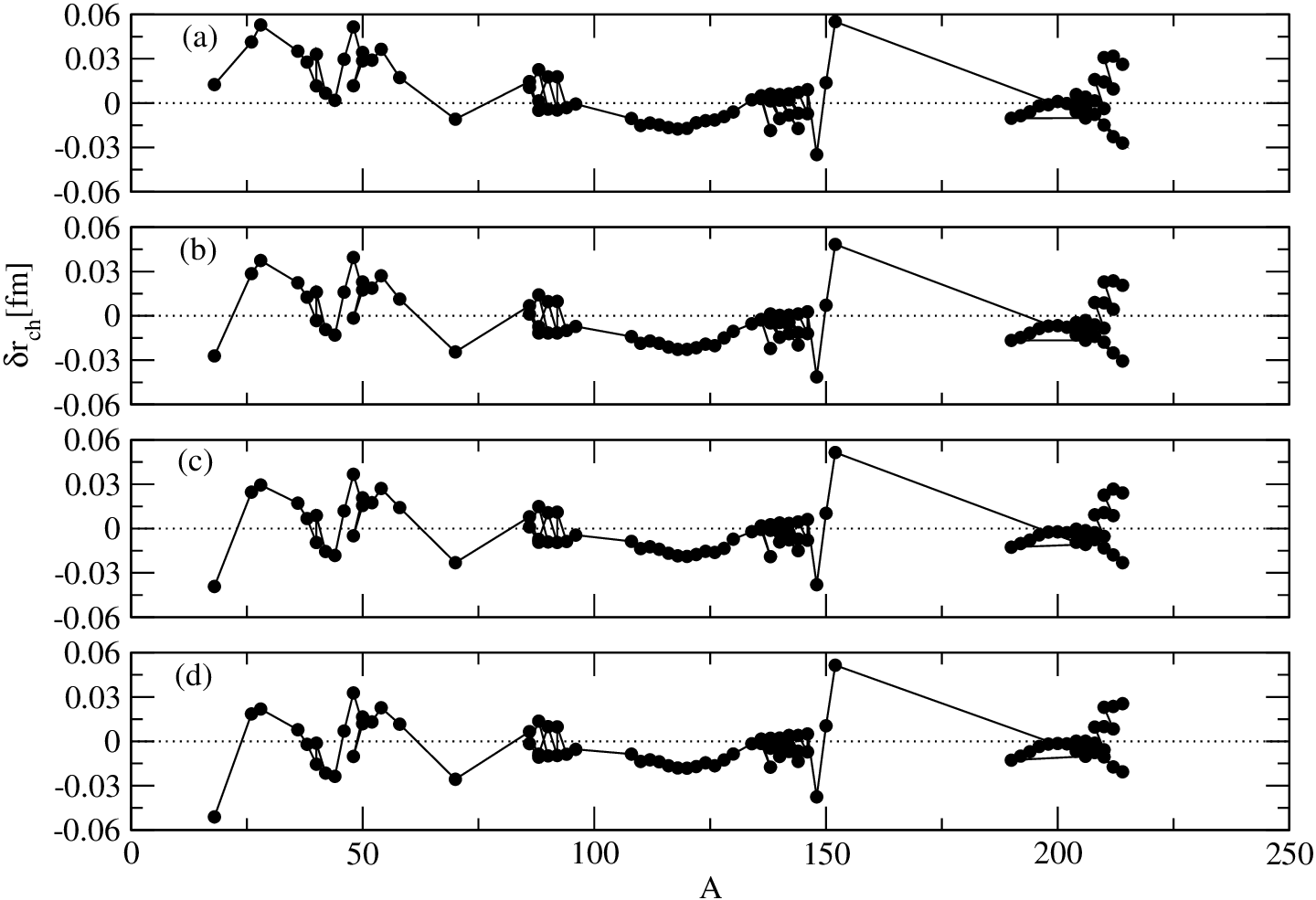}
\caption{Deviation in charge radii for 86 even-even spherical nuclei with
nucleon number between $A=16$ and $A=224$
for the EOSs corresponding to 
$\gamma$=1/6, 1/3, 1/2 and 2/3 with sets of parameters for $x_0$ in table 4 
and are shown in (a), (b), (c) and (d), respectively. 
}
\end{center}
\label{Fig. 12}
\end{figure}
%%Figure 14
\begin{figure}
\vspace{0.6cm}
\begin{center}
\includegraphics[width=0.98\columnwidth,clip=true]{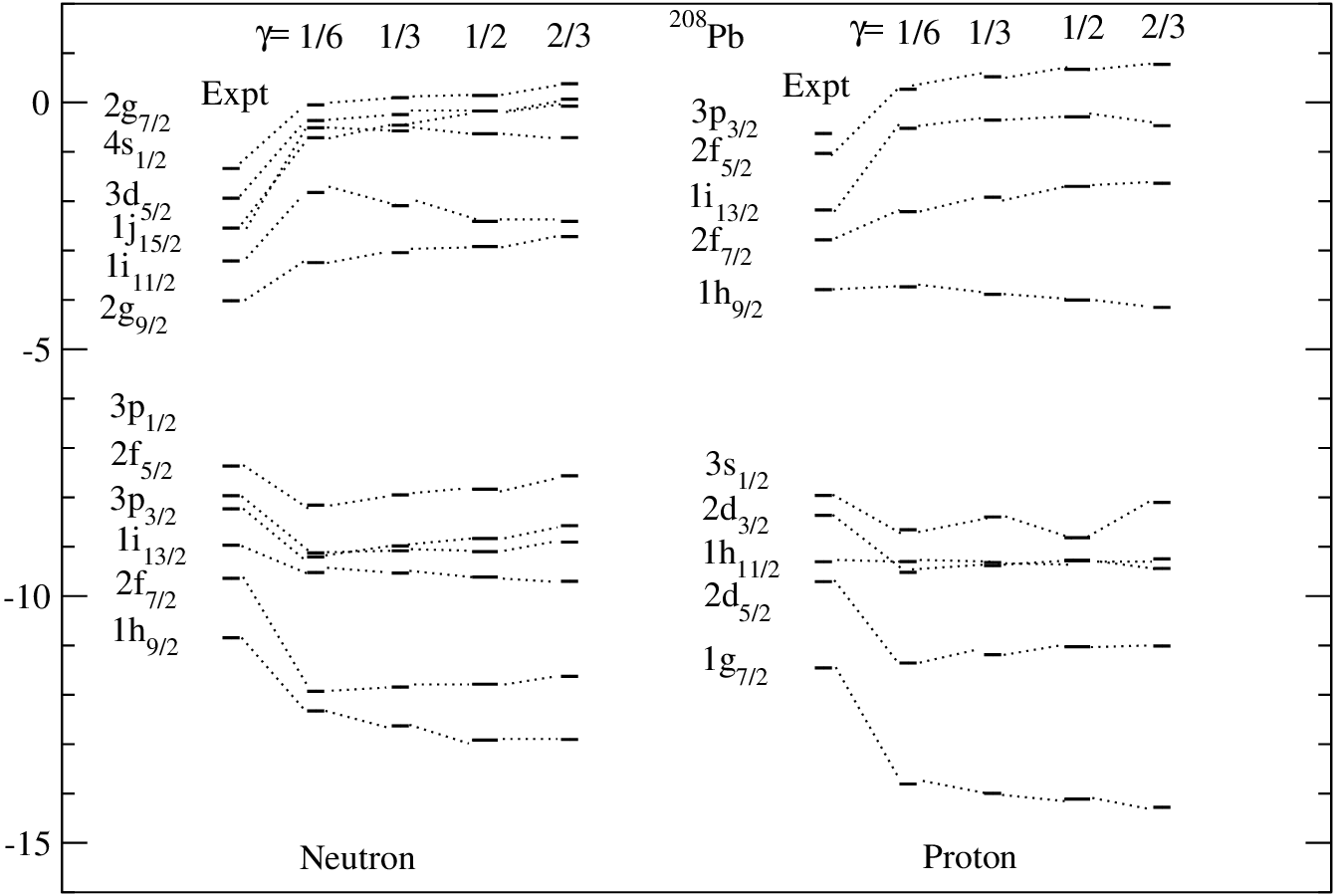}
\caption{The single particle levels for neutrons and protons in $^{208}$Pb
for the EOSs corresponding to
$\gamma$=1/6, 1/3, 1/2 and 2/3 with sets of parameters for $x_0$ in table 4
 The experimental results are also given for comparison.
}
\end{center}
\label{Fig. 13}
\end{figure}

\subsection{Finite nuclei properties with the new parameter sets}

At last we examine the ability of SEI in predicting finite nuclei properties
with the new sets of parameters where $x_0$ is fixed from the consideration
of spin polarized neutron matter as explained in section 3.1. In this new protocol, the spin-orbit strength $W_0$ is adjusted to reproduce the BE of $^{208}${Pb}.
The prediction of the deviations in BEs and charge radii for the even-even spherical
nuclei are shown in Figures 11 and 12, respectively, for the four EOSs. The {\it
rms} deviations in BE, $\delta{E_{\it rms}}$, and charge radii, $\delta{r_{\it
rms}}$, along with the values of $t_0$, $x_0$ and $W_0$ are listed in table 4.
From Figures 11 and 12 and the corresponding {\it rms} deviations given in Table 4, it can be seen that the results of finite nuclei are reproduced within reasonable accuracy for all the four EOSs. The present values of the $\delta{E_{\it rms}}$ and $\delta{r_{\it rms}}$ deviations are comparable to those obtained for the same set of nuclei with the SLy4, NL3, D1S, BCP1 and BCP2 effective forces given in Table 3 of Ref.\cite{baldo08}. 
In order to examine the predictions of single-particle levels 
and their splittings, the neutron and proton energy levels in $^{208}${Pb} are
shown in Figure 13 for the four EOSs having $\gamma$=1/6, 1/3, 1/2 and 2/3
alongwith the experimental values \cite{audi95,firestone96}. The quality of the single-particle spectra obtained in the present case is similar to that obtained
in other traditional effective forces shown in our earlier
work \cite{trr13}. From the comparison of the spectra in the four EOSs, it can be concluded that nearly similar spectra are predicted by all the four EOSs, having a tendency of widening the gaps between the single-particle levels with increase in stiffness.

\section{Summary and Conclusions}

In this work we have investigated some of the various types of nuclear matter as well as finite nuclei
with the finite range simple effective interaction. This interaction describes the N-N
force in the four basic states $SE$, $TE$, $TO$, $SO$ with the same range
but different strengths. The SEI contains in total twelve parameters. Most of these parameters are determined from the consideration of the basic properties of asymmetric nuclear matter and the momentum dependence of the mean field. In previous work \cite{trr13}, the remaining parameters---namely, $t_0$, $x_0$ and the spin-orbit strength $W_0$---were fitted to a few magic nuclei.
However, on examining the predictions of the SEI in spin polarized symmetric nuclear matter and pure neutron matter, it is found that $x_0$ and $W_0$ are 
correlated in the earlier fitting procedure that leaves some uncertainty in the values of these parameters. 
This implies that changes in the $x_0$ parameter can be compensated by small changes in $W_0$ that practically do not affect the {\it rms} deviations from the experimental binding energies and charge radii of spherical nuclei, but 
manifests in the predictions in spin polarized matter (see e.g. figure 4). 
To overcome this difficulty we connected the determination of $x_0$ to the microscopic trend of the momentum dependence of the mean field in spin polarized 
PNM. The assessment of $x_0$ was decided from the DBHF $nu$-$nd$ effective mass splitting in polarized PNM \cite{sam07} and $t_0$, determined from the binding 
energy of $^{40}$Ca, and hence acting as a connection between nuclear matter 
and the finite nucleus. 
Finally, the spin-orbit strength $W_0$ was fitted to the binding energy of $^{208}$Pb.
The parameter sets of SEI thus obtained are able to reproduce the finite nuclei results
with a similar quality to other traditional forces along with definite predictions in isospin and spin polarized matter. 

The momentum dependence of the spin-up and spin-down neutrons and the effective mas splitting in spin polarized PNM is discussed. It is shown that the spin-up neutrons have larger effective mass than the spin-down neutrons when the strength of the exchange part in the like channel is smaller in magnitude than in the unlike channel, and vice-versa. In terms of the interaction parameters, this can be interpreted from the fact that if the difference between the Wigner and Heisenberg strength parameters is attractive (repulsive)
then the single-particle potential and the effective mass for spin-up neutrons in PPNM shall be above (below) the same quantities for spin-down neutrons.
The SEI is found to be competent to reproduce the microscopic trends of the momentum dependence in NM, but requires modification in the density-dependent part for predicting the spin symmetry energy in closer agreement with the microscopic result. 
With the present density dependence of SEI the ferromagnetic and anti-ferromagnetic cases of spin polarized SNM have been examined. It is found that the SEI does not predict a ferromagnetic transition of SNM, whereas an anti-ferromagnetic transition is allowed at a density around four times the normal saturation density. This result is qualitatively similar to the result predicted by the Gogny interaction in the Fermi liquid formalism \cite{isa04}. Similar results are also observed in the case of SLy Skyrme force sets. However, the SkI3 Skyrme set allows both ferromagnetic and anti-ferromagnetic transitions in SNM, but favouring the former 
at a relatively small density in comparison to 
the latter. The predictions of SEI are analysed analytically and it is found that stability against a ferromagnetic transition requires the interaction in the $TO$ state to be repulsive, opposite to the requirement of an attractive $TO$ state interaction for stability against an anti-ferromagnetic transition. 

The simultaneous study of nuclear matter of different types and finite nuclei with the SEI practically requires standard values of the three NM properties $\rho_0$, $e(\rho_0)$ and $E_{s}(\rho_0)$. From the global description of energies and radii of finite nuclei with different sets of EOSs of SEI, the influence of the incompressibility $K(\rho_0)$ stands out in terms of the variations in the values of $T_{f_0}$ and $E_s(\rho_0)$ 
given in table 2. As the incompressibility $K(\rho_0)$ varies from 207 MeV to 
263 MeV, the Fermi kinetic energy $T_{f_0}$ decreases from 37.2 MeV to 36.1 MeV and the symmetry energy $E_s(\rho_0)$ 
decreases from 36 MeV to 35 MeV in order to reproduce the results of BEs
and charge radii with minimun {\it rms} deviations.
Thus, the SEI with the parametrization determined in this work 
can be used in the study of asymmetric NM and spin polarized NM as well as in finite nuclei. 
We have restricted our analysis of finite nuclei to even-even spherical systems. 
It is therefore important to extend this set of nuclei by including deformed nuclei. 
Work to adapt our numerical codes for deformed calculations is being undertaken and will be 
the subject of future communications.

\section*{Acknowledgments}
One author (TRR) thanks the Departament d'Estructura i Constituents de la Mat\`eria,
University of Barcelona, Spain for hospitality during the visit. The work is
covered under SAP programme of School of Physics, Sambalpur University, India.
X.V. and M.C. acknowledge partial support from the Spanish Consolider-Ingenio 2010
Programme CPAN CSD2007-00042, Grant No.\ FIS2011-24154 from MICINN and FEDER (Spain),
Grants No.\ 2009SGR-1289 and 2014SGR-401 from Generalitat de Catalunya, and the ``NewCompStar'' COST
Action MP1304.


\section*{References}


\begin{thebibliography}{99}
\bibitem{ter87}
Haar B ter and Malfliet R 1987 {\it Phys. Rep.} {\bf 149} 207
\bibitem{muth00}
Muther H and Polls A 2000 {\it Prog. Part. Nucl. Phys.} {\bf 45} 243
\bibitem{hoffma01}
Hoffmann F, Keil C M and Lenske H 2001 {\it Phys. Rev. C} {\bf 64} 034314
\bibitem{samma10}
Sammarruca F 2010 {\it Int. J. Mod. Phys. E} {\bf 19} 1259
%\bibitem{samma20}
%Alonso D and Sammarruca F 2003 {\it Phys. Rev. C} {\bf 67} 054301
%\bibitem{dbhf1}
%Anastasio M R, Celenza L S, Pong W S, and Shakin C M 1983 {\it Phys. Rep.} {\bf 100} 327
%\bibitem{dbhf2}
%Horowitz C J and Serot B D 1984 {\it Phys. Lett. B} {\bf 137} 287; {\it Nucl. Phys. A} {\bf 464} 613
%\bibitem{dbhf3}
%Brockmann R and Machleidt R 1984 {\it Phys. Lett. B} {\bf 149} 283; 1990 {\it Phys. Rev. C} {\bf 42} 1965
\bibitem{dbhf3}
Brockmann R and Machleidt R 1990 {\it Phys. Rev. C} {\bf 42} 1965
%\bibitem{ebhf04}
%Van Dalen E N E, Fuchs C and Faessler A 2004 {\it Nucl. Phys. A} {\bf 744} 227
\bibitem{ebhf05}
Van Dalen E N E, Fuchs C and Faessler A 2005 {\it Phys. Rev. Lett.}
{\bf 95} 022302
%\bibitem{bhf1}
%Brueckner K A, Levinson C A and Mahmoud H M 1954 {\it Phys. Rev.} {\bf 95} 217
%\bibitem{brueck67}
%Brueckner K A, Coon S A and Dabrowski J 1967 {\it Phys. Rev.} {\bf 168} 1184
%\bibitem{bhf2}
%Bethe H A 1956 {\it Phys. Rev.} {\bf 103} 1353
%\bibitem{bhf3}
%Goldstone J 1957 {\it Proc. R. Soc. London, Ser. A} {\bf 239} 267
%\bibitem{bhf4}
%Bethe H A 1971 {\it Annu. Rev. Nucl. Sci.} {\bf 21} 98
\bibitem{pand81}
Friedman B and Pandharipande V R 1981 {\it Nucl. Phys.} {\bf A361} 502
%\bibitem{lag81}
%Lagaris I E and Pandharipande V R 1981 {\it Nucl. Phys.} {\bf A369} 470
\bibitem{Bomb91}
Bombaci I and Lombardo U 1991 {\it Phys. Rev. C} {\bf 44} 1892
\bibitem{Wu07}
Xu J, Chen L W, Li  B A and Ma H R 2007 {\it Phys. Rev. C} {\bf 75} 014607
\bibitem{bal04}
Baldo M, Maieron C, Schuck P and Vi\~nas X 2004 {\it Nucl. Phys. A} {\bf 736}
241
\bibitem{akmal01}
Akmal A, Pandharipande V R and Ravenhall D G 1998 {\it Phys. Rev C}
{\bf 58} 1804
\bibitem{Wiringa}
%Wiringa R B 1988 {\it Phys. Rev. C} {\bf 38} 2967;
Wiringa R B, Fiks V, Fabrocini A 1988 {\it Phys. Rev. C} {\bf 38} 1010
%
%\bibitem{boguta77}
%Boguta J, Bodmer A R 1977 {\it Nucl. Phys. A} {\bf 292} 413
%\bibitem{sero86}
%Serot B D and Walecka J D 1986 {\it Adv. Nucl. Phys.} {\bf 16} 1
%\bibitem{rein89}
%Reinhard P G 1989 {\it Rep. Prog. Phys.} {\bf 52} 439
\bibitem{ring96}
Ring P 1996 {\it Prog. Part. Nucl. Phys.} {\bf 37} 193
\bibitem{brink72}
Vautherin D and Brink D M 1972 {\it Phys. Rev.} {\bf C 5} 626
\bibitem{brack85}
Brack M, Guet C and Hakansson H -B 1985 {\it Phys. Rep.} {\bf 123} 275
%\bibitem{trein86}
%Treiner J {\it et al} 1986 {\it Ann. Phys. (N.Y.)} {\bf 170} 406
\bibitem{trr98}
Behera B, Routray T R and Satpathy R K 1998 {\it J. Phys. G: Nucl. Part. Phys.}
{\bf 24} 2073
\bibitem{ston07}
Stone J R and Reinhard P -G 2007 {\it Prog. Part. Nucl. Phys.} {\bf 58} 587
%
%\bibitem{ring90}
%Gambhir Y K, Ring P and Thimet A 1990 {\it Ann. Phys.} (N.Y.) {\bf 198} 132
\bibitem{lala97}
Lalazissis G A, Konig K and Ring P 1997 {\it Phys. Rev. C} {\bf 55} 540
\bibitem{patra01}
Del Estal M, Centelles M, Vi\~nas X and Patra S K 2001
{\it Phys. Rev. C} {\bf 63} 024314
%\bibitem{nl3im}
%Lalazissis G A, Karatzikos S, Fossion R, Pena Arteaga D, Afanasjev A V
%and Ring P 2009 {\it Phys. Lett. B} {\bf 671} 36
%\bibitem{ddme1}
%Niksic D, Vretenar D, Finelli P and Ring P 2002 {\it Phys. Rev. C} {\bf 66} 024306
\bibitem{ddme2}
Lalazissis G A, Niksic T, Vretenar D and Ring P 2005 {\it Phys. Rev. C} {\bf 71} 024312
\bibitem{todd05}
Todd-Rutel B G and Piekarewicz J 2005 {\it Phys. Rev. Lett.} {\bf 95} 122501
\bibitem{klahn06}
Klahn T et {\it al.} 2006 {\it Phys. Rev. C} {\bf 74} 035802
\bibitem{roca11}
Roca-Maza X, Vi\~nas X, Centelles M, Ring P and Schuck P
2011 {\it Phys. Rev. C} {\bf 84} 054309
\bibitem{afan13}
Afanasjev A V, Agbemava S E, Ray D and Ring P 2013 {\it Phys. Lett. B} {\bf 726} 680
%
%\bibitem{sk59}
%Skyrme T H R 1959 {\it Nucl. Phys.} {\bf 9} 615
\bibitem{bein75}
Beiner M, Flocard H, Giai Nguyen Van and Quentin P 1975 {\it Nucl. Phys. A} {
\bf 238} 29
%\bibitem{koh76}
%Kohler H S 1976 {\it Nucl. Phys. A} {\bf 258} 301
%\bibitem{kriv80}
%Krivine H, Treiner J and Bohigas O 1980 {\it Nucl. Phys. A} {\bf 336} 155
\bibitem{rein95}
Reinhard P -G and Flocard H 1995 {\it Nucl. Phys. A} {\bf 584} 467
\bibitem{chab97}
Chabanat E, Bonche P, Hansel P, Meyer J and Schaeffer R 1997 {\it Nucl.
Phys. A} {\bf 627} 710; 1998 {\it Nucl. Phys. A} {\bf 635} 231
%\bibitem{chab98}
%Chabanat E, Bonche P, Hansel P, Meyer J and Schaeffer R 1998 {\it Nucl. Phys. A} {\bf 635} 231
\bibitem{gor10}
Goriely S, Chamel N and Pearson J M 2010 {\it Phys. Rev. C} {\bf 82} 035804
\bibitem{erl12}
Kortelainen M {\it et al} 2014 {\it Phys. Rev. C} {\bf 89} 054314 
%
\bibitem{gogny80}
Decharge J and Gogny D 1980 {\it Phys. Rev. C} {\bf 21} 1568
\bibitem{blaizot95}
Blaizot J P, Berger J F, Decharge J and Girod M 1995
{\it Nucl. Phys. A} {\bf 591} 435
\bibitem{chappert08}
Chappert F, Girod M and Hilaire S 2008 {\it Phys. Lett. B} {\bf 668} 420
\bibitem{goriely09}
Goriely S, Hilaire S and Girod M 2009 {\it Phys. Rev. Lett.} {\bf 102} 242501
%\bibitem{anant83}
%Anantaraman N, Toki H and Bertsch G 1983 {\it Nucl. Phys. A}{\bf 398} 279
%\bibitem{hof98}
%Hofmann F and Lenske H 1998 {\it Phys. Rev. C} {\bf 57} 2281
\bibitem{nakada03}
Nakada H 2003 {\it Phys. Rev. C} {\bf 68} 014316; 2008 {\it Phys. Rev. C} {\bf 78} 054301
\bibitem{than09}
Than H S, Khoa Dao T, and Giai Van N 2009 {\it Phys. Rev. C} {\bf 80}, 064312
%
\bibitem{trr02}
Behera B, Routray T R, Sahoo B and Satpathy R K 2002 {\it Nucl. Phys. A}
{\bf 699} 770
%
\bibitem{trr13}
Behera B, Vi\~nas X, Bhuyan M, Routray T R, Sharma B K and Patra S K 2013 {\it J. Phys G: Nucl. Part. Phys.} {\bf 40} 095105
%
\bibitem{raim14}
Raimondi F, Bennaceur K and Dobaczewski J 2014 {\it J. Phys G: Nucl. Part. Phys.} {\bf 41} 055112 
%
\bibitem{ma04}
Ma Z Y, Rong J, Chen B Q, Zhu Z Y and Song H Q 2004 {\it Phys. Lett.}
{\bf B604} 170
\bibitem{sammu05}
Sammarruca F, Barredo W and Krastev P 2005 {\it Phys. Rev. C}
{\bf 71} 064306
\bibitem{zuo05}
Zuo W, Gao L G, Li B A, Lombardo U and Shen C W 2005 {\it Phys. Rev. C}
{\bf 72} 014005
%
\bibitem{kubis97}
Kubis S, Kutschera M 1997 {\it Phys. Lett. B} {\bf 399} 191
%\bibitem{greco01}
%Greco V, Matera M, Colonna M, Di Toro M, Fabbri G 2001 {\it Phys. Rev. C} {\bf 63} 035202
%\bibitem{greco1a}
%Greco V, Colonna M, Di Toro M and Fabbri G 2001 {\it Phys. Rev. C} {\bf 63} 045203
\bibitem{greco03}
Greco V, Baran V, Colonna M, Di Toro M, Gaitanos and Wolter H H
2003 {\it Phys. Lett. B} {\bf 562} 215
%\bibitem{ebhf07}
%Van Dalen E N E, Fuchs C and Faessler A 2007 {\it Eur. Phys. J. A} {\bf 31} 29 
%
\bibitem{lane62}
Lane A M 1962 {\it Nucl. Phys.} {\bf 35} 676
\bibitem{hod94}
Hodgson P E 1994 {\it The Nucleon Optical Model} (Singapore:World Scientific)
p 613
%
\bibitem{dutra12}
Dutra M, Lourenco O, Martins J S S, Delfino A, Stone J R and Stevenson P D 2012 {\it Phys. Rev. C} {\bf 85} 035201
%
\bibitem{behera05}
Behera B, Routray T R and Pradhan A 2005 {\it Mod. Phys. Lett. A}
{\bf 20} 2639
%
\bibitem{gale87}
Gale C, Bertsch G F, Das Gupta S 1987 {\it Phys. Rev. C} {\bf 35} 1666
\bibitem{bers88}
Bertsch G F, Das Gupta S 1988 {\it Phys. Rep.} {\bf 160} 189
\bibitem{welke88}
Welke G M, Prakash M, Kuo T T S, Das Gupta S, Gale C,
1988 {\it Phys. Rev. C} {\bf 38} 2101
%
\bibitem{gale90}
Gale C, Welke G M, Prakash M, Lee S J, Das Gupta S,
1990 {\it Phys. Rev. C} {\bf 41} 1545
%ffmann, C. M. Keil and H. Lenske, Phys. Rev. C {\bf 64}, 034314 (2001).
\bibitem{cser92}
Csernai L P, Fai G, Gale C and Osnes E 1992 {\it Phys. Rev. C}
{\bf 46} 736
\bibitem{pan93}
Pan Q, Danielewicz P 1993 {\it Phys. Rev. Lett.} {\bf 70} 2062
\bibitem{zhang94}
Zhang J, Das Gupta S, Gale C 1994 {\it Phys. Rev. C} {\bf 50} 1617
\bibitem{Danielz00}
Danielewicz P 2000 {\it Nucl. Phys. A} {\bf 673} 375
%
%\bibitem{behera05}
%Behera B, Routray T R and Pradhan A 2005 {\it Mod. Phys. Lett. A}
%{\bf 20} 2639 
\bibitem{trr05}
Behera B, Routray T R, Pradhan A, Patra S K and Sahu P K 2005
{\it Nucl. Phys. A} {\bf 753}, 367.
\bibitem{trr07}
Behera B, Routray T R, Pradhan A, Patra S K and Sahu P K 2007
{\it Nucl. Phys. A} {\bf 794}, 132.
\bibitem{trr09}
Behera B, Routray T R and Tripathy S K 2009 {\it J. Phys. G:
Nucl. Part. Phys.} {\bf 36}, 125105.
\bibitem{trr11}
Behera B, Routray T R and Tripathy S K 2011 {\it J. Phys. G:
Nucl. Part. Phys.} {\bf 38}, 115104.
%
\bibitem{baldo08}
Baldo M, Schuck P and Vi\~nas X 2008 {\it Phys. Lett. B} {\bf 663} 390
%
\bibitem{trr97}
Behera B, Routray T R and Satpathy R K 1997 {\it J. Phys G: Nucl.
Part Phys.} {\bf 23} 445
%
\bibitem{Danielz02}
Danielewicz P, Lacey R and Lynch W G 2002 {\it Science} {\bf 298} 1592
%
\bibitem{pop06}
Popov S, Grigorian H, Turolla R and Blaschke D 2006 {\it Astron. Astrophys.} {\bf 448} 327
\bibitem{blas04}
Blaschke D, Grigorian H and Voskrenensky D 2004 {\it Astron. Astrophys.} {\bf 424} 979
%
\bibitem{vinas00}
Soubbotin V B and Vi\~nas X 2000 {\it Nucl. Phys. A} {\bf 665} 291
\bibitem{vinas03}
Soubbotin V B, Tselyaev V I and  Vi\~nas X  2003
{\it Phys. Rev. C} {\bf 67} 014324
%
\bibitem{ber91}
Bertsch G F and Esbensen H 1991 {\it Ann. Phys.} {\bf 209} 327
%
\bibitem{vid84}
Vidaurre A, Navarro J and Bernabeu J 1984 {\it Astron. Astrophys.} {\bf 135}
361
\bibitem{kuts94}
Kutschera M and W\'ojcik W 1994 {\it Phys. Lett. B} {\bf 325} 271
\bibitem{pandh72}
Pandharipande V R, Garde V K and Srivastava J K 1972 {\it Phys. Lett. B} {\bf 38} 485
\bibitem{uma97}
Uma Maheswari V S, Basu D N, De J N and Samaddar S K 1997 {\it Nucl. Phys. A} {\bf 615} 516 
\bibitem{fan01}
Fantoni S, Sarsa A and Schmidt E 2001 {\it Phys. Rev. Lett.} {\bf 87} 181101
\bibitem{vid02}
Vida\~na I, Polls A and Ramos A 2002 {\it Phys. Rev. C} {\bf 65} 035804
\bibitem{bom02}
Vida\~na I and Bombaci I 2002 {\it Phys. Rev. C} {\bf 66} 045801
\bibitem{sam07}
Sammarruca F and Krastev P G 2007 {\it Phys. Rev. C} {\bf 75} 034315
\bibitem{sam10}
Sammarruca F 2010 {\it Phys. Rev. C} {\bf 82} 027307
\bibitem{sam11}
Sammarruca F 2011 {\it Phys. Rev. C} {\bf 83} 064304
\bibitem{zuo03}
%Zuo W, Lombardo U and Shen C W {\it arXiv:nucl-th/0204056};
Zuo W, Shen C W and Lombardo U 2003 {\it Phys. Rev. C} {\bf 67} 037301
\bibitem{isa04}
Isayev A A and Yang J 2004 {\it Phys. Rev. C} {\bf 70} 064310
\bibitem{isa204}
Isayev A A and Yang J 2004 {\it Phys. Rev. C} {\bf 69} 0025801
\bibitem{isa06}
Isayev A A 2006 {\it Phys. Rev. C} {\bf 74} 057301
\bibitem{bod07}
Bordbar G H and Bigdeli M 2007 {\it Phys. Rev. C} {\bf 76} 035803
\bibitem{big10}
Bigdeli M 2010 {\it Phys. Rev. C} {\bf 82} 054312
%
\bibitem{rio05}
Rios A,  Polls A and Vida\~na I 2005 {\it Phys. Rev. C} {\bf 71} 055802
%
\bibitem{audi95}
Audi G and Wapstra A H 1995 {\it Nucl. Phys. A} {\bf 595} 409
\bibitem{firestone96}
Firestone R B {\it et al.}, {\it Table of Isotopes}, {\bf 8th edition} (1996)
(John Wiley $\&$ Sons, New York).
\end{thebibliography}
\end{document}